\documentclass[12pt]{iopart}
\usepackage{graphicx}
\usepackage{color}
\newcommand{\be}{\begin{equation}}
\newcommand{\ee}{\end{equation}}
\newcommand{\ba}{\begin{eqnarray}}
\newcommand{\ea}{\end{eqnarray}}
\newcommand{\mpl}{m_{\mathrm{pl}}}

\newcommand{\ltsim}{\protect\raisebox{-0.5ex}
  {$\:\stackrel{\textstyle <}{\sim}\:$}}
\newcommand{\gtsim}{\protect\raisebox{-0.5ex}
  {$\:\stackrel{\textstyle >}{\sim}\:$}}

\begin{document}

\title[Probing polarization states of primordial gravitational waves]
{Probing polarization states of primordial gravitational waves with CMB 
anisotropies}

\author{Shun Saito$^1$, Kiyotomo Ichiki$^2$ and Atsushi Taruya$^2$}
\address{$^1$ Department of Physics, The University of Tokyo, 
  Tokyo 113-0033, Japan}
\address{$^2$ Research Center for the Early Universe (RESCEU),
Graduate School of Science, The University of Tokyo, Tokyo 113-0033, Japan}
\eads{\mailto{ssaito@utap.phys.s.u-tokyo.ac.jp}
}

\begin{abstract}
We discuss the polarization signature of primordial gravitational 
waves imprinted in cosmic microwave background (CMB) anisotropies. The 
high-energy physics motivated by superstring theory or M-theory generically 
yields parity violating terms, which may produce a circularly polarized 
gravitational wave background (GWB) during inflation. 
In contrast to the standard prediction of inflation with un-polarized 
GWB, circularly polarized GWB generates non-vanishing TB and EB-mode 
power spectra of CMB anisotropies. We evaluate the TB and EB-mode power 
spectra taking into account the secondary anisotropies induced by 
the reionization and investigate the dependence 
of cosmological parameters. We then discuss current constraints on the 
circularly 
polarized GWB from large angular scales ($\ell\le 16$) of the three year WMAP 
data. Prospects for future CMB experiments are also investigated based 
on a Monte Carlo analysis of parameter estimation, showing that the 
circular polarization degree, $\varepsilon$, which is 
the asymmetry of the tensor power spectra between right- and 
left-handed modes normalized by the total amplitude, 
can be measured down to $|\varepsilon|\gtsim0.35(r/0.05)^{-0.6}$. 
\end{abstract}

\maketitle

\section{Introduction}
\label{sec:intro}

 The gravitational wave background (GWB) originating from inflation provides a 
direct probe of inflation against which we can test inflationary models. 
In particular, the energy scale of inflation can be estimated from the amplitude
 of the GWB. Combining with observations of scalar-type fluctuations, 
 the detection of the GWB directly 
constrains the inflaton potential \cite{Bock:2006yf}. Currently, there is no 
rigorous constraint on the amplitude of the GWB characterized by the tensor-to-scalar 
ratio, $r\equiv \Delta^{2}_{\rm gw}(k_{0})/
\Delta^{2}_{\mathcal{R}}(k_{0})$\footnote{In this 
paper, we adopt the conventional value $k_{0}=0.002\mathrm{Mpc^{-1}}$ as the 
pivot scale.}. In the standard scenario of slow-roll 
inflation, the GWB is 
expected to have a nearly scale-invariant spectrum, suggesting that the GWB would 
be detectable in a wide range of wavelengths or frequencies. For a large-scale 
experiment, polarization anisotropies of the cosmic microwave background (CMB) 
would be a powerful tool to search for primordial tensor fluctuations. 
Indeed, post-Planck experiments such as 
SPIDER\footnote{{\tt http://www.astro.caltech.edu/$\sim$lgg/spider$_-$front.htm}} 
and CMBpol (or Inflation Probe) in the Beyond Einstein program of 
NASA\footnote{{\tt http://universe.nasa.gov/program/inflation.html}} 
is dedicated to measuring the B-mode polarization anisotropies originating 
from the inflationary GWB, 
with expected precision level $r\sim 10^{-3}$ \cite{Verde:2005ff,Amblard:2006ef}. 
On the other hand, a direct measurement of the stochastic 
GWB might be possible for a small-scale experiment, especially using 
space-based laser interferometers 
\cite{Cooray:2005xr,Smith:2005mm,Smith:2006xf,Kudoh:2005a}. 
Proposed missions such as the 
Big-Bang Observer (BBO) \cite{BBO} and the deci-hertz 
gravitational-wave observatory (DECIGO) 
\cite{Seto:2001qf,2006CQGra..23S.125K} indeed aim 
at detecting the primordial gravitational waves at the frequencies 
$f\sim 0.1-1$Hz. 
Notice that the observational frequencies (or wavelengths) of the space 
interferometers are greatly different from those of the CMB experiments by 
$16$ orders of magnitude.  
This implies that the combination of both experiments provides a stringent 
constraint on the dynamics of inflation. 
\par

Meanwhile, motivated by high-energy physics, there are numerous 
discussions on the corrections to the prediction of standard inflationary 
models. For example, some inflationary models contain parity-violating 
interaction terms, as generic predictions 
of superstring theory/ M-theory \cite{Green84,Witten84}. Among these, the 
Chern-Simons term, which is a higher-order curvature term coupled to the scalar 
field, appears through the Green-Schwarz mechanism and the cosmological 
implications of  the Chern-Simons terms has been extensively discussed 
\cite{Lue:1998mq,Choi:1999zy,Jackiw:2003pm,Lyth:2005jf,Alexander:2004wk,
Alexander06_1,Alexander06_2,Satoh:2007gn}. Such parity-violating terms 
directly affect the tensor-type 
perturbations \cite{Antoniadis:1993jc,Brustein:1999yq} 
and the polarization modes of the resulting tensor fluctuations 
becomes asymmetric, leading to a circularly polarized GWB \cite{Lue:1998mq,
Choi:1999zy,Jackiw:2003pm,Satoh:2007gn}. 
In this respect, the detection of a circularly 
polarized GWB would be a direct signature of the cosmological parity-violation 
and it might also imply that there is a fundamental theory of particle physics beyond 
the standard model 
\footnote{There exists another source to generate a circular polarized GWB 
in the early universe, i.e., primordial helical turbulence produced during 
a first-order phase transition \cite{Caprini:2003vc,Kahniashvili:2005qi}. However, 
the wavelength of the produced GWB is much smaller than the scale of CMB 
observations. }. Since the sensitivity of the forthcoming CMB experiments will 
be improved significantly, it is timely to explore the possibilities of 
measuring the signature of new interaction terms through the CMB anisotropies. 

\par

In this paper, we discuss in some detail, the observational possibilities of 
using CMB anisotropy measurements to 
probe the additional signature imprinted in the primordial gravitational waves. 
According to \cite{Lue:1998mq}, a circularly polarized GWB produces a 
non-trivial correlation between the temperature and the polarization 
anisotropies.  As a result, the cross power spectra between temperature 
and B-mode polarization 
becomes non-vanishing. They calculated the TB-mode spectrum 
in an idealistic situation with a large tensor-to-scalar ratio,  
neglecting the secondary anisotropies.  
In the present paper, extending their analysis, we quantitatively evaluate the 
TB mode spectrum taking into account the effect of secondary anisotropies.  
Also, we calculate another non-vanishing spectrum, the EB-mode spectrum.  
Based on the three year data of Wilkinson Microwave Anisotropy Probe (WMAP) 
\cite{Spergel:2006hy,Page:2006hz}, we discuss the current constraint 
on the degree of polarization of the primordial GWB. Further, we address 
future prospects for the PLANCK satellite or cosmic-variance limited experiments 
and estimate the extent to which the degree of polarization is constrained from 
future observations. 
\par

This paper is organized as follows. In \S \ref{sec2}, we briefly 
describe a mechanism to generate circular polarization of the GWB through 
the gravitational Chern-Simons term. 
In \S \ref{sec3}, the influence of the polarized GWB on the CMB anisotropies is 
discussed. 
Based on this, the CMB power spectra originating from circularly polarized 
GWB are 
calculated in \S \ref{sec4}. The current constraints and future prospects are 
discussed in \S \ref{sec5}. 
Finally, \S \ref{sec6} is devoted to discussion and conclusions. 


\section{Polarized gravitational waves from gravitational 
Chern-Simons term} 
\label{sec2}

 In this section, we briefly review a mechanism to generate a circularly 
polarized GWB by the gravitational Chern-Simons (gCS) 
term \cite{Lue:1998mq}. In superstring 
theory/M-theory, there exist scalar fields coupled with anti-symmetric tensor 
$F \wedge F\equiv \epsilon^{\alpha\beta\gamma\delta}F_{\alpha\beta}
F_{\gamma\delta}$ and/or $R\wedge R\equiv \epsilon^{\alpha\beta\gamma\delta}
R_{\alpha\beta}^{\ \ \mu\nu}R_{\gamma\delta\mu\nu}$, where $F_{\mu\nu}$ 
is the field strength of the electromagnetic field, $R_{\alpha\beta\gamma\delta}$ 
is the Riemann tensor, and $\epsilon^{\alpha\beta\gamma\delta}$ is a totally 
antisymmetric Levi-Civita tensor 
density \cite{Green84,Witten84}. These two terms are referred to as the 
electromagnetic and the gravitational Chern-Simons term, respectively. 
The presence of such parity-violating terms plays an important role for  
several cosmological issues 
such as structure formation involving axions \cite{Choi:1999zy} and 
leptogenesis or baryogenesis in the early 
universe \cite{Lyth:2005jf,Alexander:2004wk,
Alexander06_1,Alexander06_2}. In a homogeneous and isotropic 
background spacetime, the electromagnetic Chern-Simons term affects neither 
the evolution of the background spacetime nor the evolution of fluctuations 
as long as only linear perturbation is considered 
\cite{Brustein:1998du}. Therefore, we consider the gCS term only:
\be
 S_{CS}=\frac{\mpl^{2}}{64\pi}\int d^{4}x f(\phi)R\wedge R\ ,
\label{eq:gCS}
\ee
 where $\mpl$ is the Planck mass, and the function $f(\phi)$ represents a generic 
coupling to the scalar field $\phi$. In some cases, the scalar field $\phi$ is 
identified with the inflaton field during inflation. 
As long as the inflaton field $\phi$ is homogeneous and constant in time, equation 
(\ref{eq:gCS}) is just a surface term, and it does not contribute at all to 
classical gravitational dynamics. Thus, after the end of inflation, we expect 
that the classical evolution without gCS term is recovered and no anomalous 
parity violation appears. 
Moreover, the gCS term also does not affect the evolution of the background and 
scalar perturbations in the linear regime \cite{Antoniadis:1993jc,Brustein:1999yq}. 
Thus, if we ignore the vector perturbation, the influence of the gCS term only 
appears in the evolution of tensor perturbations.

Let us linearize the Einstein-Hilbert action in the presence of the gCS term. Assuming 
a flat Friedmann-Robertson-Walker cosmology, the corresponding metric neglecting the 
scalar perturbation takes the following form: 
\be
 ds^{2}=a^{2}(\eta)[-d\eta^{2}+(\delta_{ij}+h_{ij})dx^{i}dx^{j}]\ ,\\
\ee
with $h_{ij}$ being a transverse and traceless tensor, i.e., 
$\partial^{j}h_{ij}=h^{i}_{\ i}=0$. Expanding the action up to the second order in 
the gravitational wave tensor $h_{ij}$,  the evolution equation for tensor 
fluctuations is obtained in Fourier space as \cite{Alexander:2004wk}
\be
 (\mu^{s})''+\left(k^{2}-\frac{{z^{s}}''}{z^{s}}\right)\mu^{s}=0\ ,
\label{eq:equation_of_motion00}
\ee
 where the subscript $'$ denotes the derivative with respect to $\eta$, 
 the amplitude $\mu^{s}(\eta)$ is defined 
by $\mu^{s}(\eta)\equiv z^{s}h^{s}$, and the variable $z^{s}$ is defined 
by
\ba
 && z^{s}(\eta,{\bf k}) \equiv a(\eta)\sqrt{1-\lambda^{s}k\frac{f'}{a^{2}}}\ ,
\\
 && \left\{\begin{array}{c}
 \lambda^{\rm R}=+1\\
 \lambda^{\rm L}=-1\end{array}\right.\ ,
\ea
 where subscript $s$ stands for a circularly polarized state, $s=$R, L. We define the 
right-handed or left-handed circular polarized state by its helicity:
\ba
 &&h_{ij}(\eta,{\bf x})=\frac{1}{(2\pi)^{3/2}}\int d{\bf k} 
\sum_{s={\rm R,L}}e^{s}_{ij}({\bf k})h^{s}({\bf k})e^{i{\bf k}\cdot{\bf x}}\ ,\\
 &&ik_{c}\epsilon_{a}^{\ cd}e^{\rm R}_{bd}  =  ke^{\rm R}_{ab}\ ,
\label{eq:def_R-handed}
\\
 &&ik_{c}\epsilon_{a}^{\ cd}e^{\rm L}_{bd}  =  -ke^{\rm L}_{ab}\ ,
\label{eq:def_L-handed}
\ea
where $e^{\rm R,L}_{ij}$ is the polarization tensor for right-handed or left-handed 
circular polarization state. 
\par

In equation (\ref{eq:equation_of_motion00}),  
the important point is that the term ${z^{s}}''/z^{s}$ 
depends not only on time, but also on the polarization mode. This readily 
implies that asymmetry of the amplitude in left- and right-handed modes 
may be produced, leading to 
a circularly polarized GWB. Apart from the helicity-dependent nature, the evolution 
equation (\ref{eq:equation_of_motion00}) is a standard form of the harmonic 
oscillator and one may address the quantum-mechanical generation of the GWB as 
in the case of simple inflation models. However, 
there exists a subtle issue on the break-down of linear theory arising from 
the singularity of the effective potential. 
Although the analysis under tractable conditions shows that 
the produced polarization-degree of the primordial fluctuations will be small 
\cite{Alexander:2004wk},  the result might not be appropriate for the practical 
cases. The quantitative prediction of the primordial spectrum may be a serious problem 
in the predictability of the inflation model. We do not discuss in details 
the primordial spectrum of circular polarized GWBs, but rather, we focus on the 
detectability of primordial circularly polarized GWBs.

\section{CMB power spectra from circular polarization of the GWB} 
\label{sec3}

In this section, we discuss CMB anisotropies originated from 
a circular-polarized GWB. 
While we particularly consider the circularly polarized GWB, 
the linearly polarized GWB is shown to have no effect on the 
CMB power spectra due to the symmetry associated with 
spin nature of linearly polarized gravitational waves. 
The details are discussed in \ref{sec:linear-polarized}.\par

Similar to scalar-type fluctuations, tensor-type fluctuations 
(i.e., GWB) cause a small perturbation in the photon path, producing 
CMB temperature and polarization anisotropies 
\cite{Zaldarriaga:1996xe,Kamionkowski:1996ks,Hu:1997hp}. 
For a temperature fluctuation map $T(\hat{n})$ as a function of sky position   
$\hat{n}$, let us expand it in the spherical harmonics, 
$Y_{\ell m}(\hat{n})$. 
We denote the expansion coefficients by $a^{\rm T}_{\ell m}$. Furthermore, 
polarization maps for the Stokes 
parameters $Q(\hat{n})$ and $U(\hat{n})$, which characterize the linear 
polarization state of the CMB, are obtained and are expanded by the 
spin-weighted harmonics $Y^{\pm 2}_{\ell m}(\hat{n})$. The coefficients of 
these polarization anisotropies are decomposed into an electric part, 
$a^{\rm E}_{\ell m}$, 
and a magnetic part, $a^{\rm B}_{\ell m}$ \cite{Zaldarriaga:1996xe}. 
Apart from a tiny non-Gaussianity, 
these coefficients $a^{\rm X}_{\ell m}$ (X=T,E,B) are statistically 
characterized by 
Gaussian statistics with zero mean. In the case of the 
two-point statistics of CMB temperature and polarization anisotropies are 
completely specified by the six (TT, EE, BB, TE, TB ,EB) power spectra defined 
as the rotationally-invariant quantities: 
\be
 C^{\rm XX'}_{\ell} \equiv \frac{1}{2\ell+1}\sum_{m}\left\langle 
\frac{(a^{X*}_{\ell m}a^{\rm X'}_{\ell m}+a^{\rm X}_{\ell m}a^{\rm X'*}_{\ell m})}{2}
\right\rangle\ , 
\label{eq:definition of CMB power spectra}
\ee
in terms of which,
\be
 \langle a^{\rm X*}_{\ell'm'}a^{\rm X'}_{\ell m}\rangle = 
C^{\rm XX'}_{\ell}\delta_{\ell\ell'}\delta_{mm'}\ ,
\ee
 where X and X$'$ stand for T,E and B.

Usually, the tensor perturbation produces both 
EE- and BB-mode polarization power spectra, but cross power spectra of TB- and 
EB-modes should vanish because of the parity conservation of the perturbations. 
However, a circularly-polarized 
GWB manifestly violates the parity symmetry, leading to non-zero values of the 
TB- and EB-mode power spectra. To be more precise, we write down the relation 
between the CMB anisotropy 
power spectra and the primordial power spectra of the GWB \cite{Zaldarriaga:1996xe}: 
\ba
 C^{{\rm XX'}(t)}_{\ell}  &=&  (4\pi)^{2}\int k^{2}dk 
[P^{t{\rm L}}(k)+P^{t{\rm R}}(k)]
\Delta^{t}_{{\rm X}\ell}(k)\Delta^{t}_{{\rm X'}\ell}(k)\ ;
\ea
for XX$'$=TT, EE, BB and TE, and  
\ba
 C^{{\rm YY'}(t)}_{\ell}  &=&  
(4\pi)^{2}\int k^{2}dk [P^{t{\rm L}}(k)-P^{t{\rm R}}(k)]
\Delta^{t}_{{\rm Y}\ell}(k)\Delta^{t}_{{\rm Y'}\ell}(k)\ ;
\ea 
for YY$'$=TB and EB. 
Here, subscript $^{(t)}$ indicates the contribution from tensor mode and 
$\Delta^{t}_{{\rm X}\ell}(k)$ is  photon's transfer function for $X$ 
(see \ref{sec:power_spectra}). The quantities $P^{t\,{\rm s}}(k)$ 
($s={\rm L,\,R}$) are the primordial power spectra of GWB in terms of the circular 
polarization basis. The circularly polarized GWB implies 
$P^{t{\rm L}}(k)\neq P^{t{\rm R}}(k)$, 
which clearly yields non-zero TB- and EB-mode power spectra.

Here, to characterize the polarization degree of GWB, 
we introduce the new variable $\varepsilon$ defined by
\ba
 P^{t{\rm R}}(k) &\equiv & \frac{1}{2}\left(1+\varepsilon\right) P^{t}(k)\ ,\\
 P^{t{\rm L}}(k) &\equiv & \frac{1}{2}\left(1-\varepsilon\right) P^{t}(k)\ ,\\
 P^{t}(k) &\equiv& P^{t{\rm L}}(k)+P^{t{\rm R}}(k)\ .
\ea
The variable $(\varepsilon+1)/2$ is the fractional power of right-handed GWB with 
respect to that of total GWB. Therefore $\varepsilon$ characterizes the degree of 
parity violation. For instance, $\varepsilon=-1,~0$ and $1$ respectively indicate 
perfectly left-handed polarized, un-polarized, and perfectly right-handed polarized 
GWB. Hereafter, we simply assume that $\varepsilon$ is scale-independent,  
which might be a good approximation in the slow-roll regime \cite{Alexander:2004wk}.
\par

Notice that the TT-, EE-, BB- and TE-mode power spectra remain unchanged 
irrespective 
of the parity violation. Thus, for CMB experiments, a measurement of the 
TB- and EB-mode power spectra is a unique probe to 
search for the parity violation in the early universe. Observationally,  
TB and EB-mode spectra are often used for a consistency null check to determine 
whether or not the 
foreground contamination is removed \cite{Page:2006hz}. However, in our case, 
the non-vanishing values of the TB and EB-modes are essential. In this sense, 
detection of a circularly polarized GWB should be carefully investigated in practice, 
since the incomplete foreground subtraction 
may lead to a false detection. Nevertheless, in the next section, 
we will show that TB- and EB-mode 
power spectra originating from the circularly polarized GWB have some characteristic 
features, 
especially on large-angular scales, which might be helpful to discriminate 
the primordial origin from foreground contamination. 
Moreover, note that the signals of TB- and EB-modes power spectra originating 
circularly polarized GWB do not depend on the wavelength of CMB photon in contrast to 
some foreground sources.
\par
Finally, we comment on the TB- and EB-mode power spectra generated through 
the electromagnetic Chern-Simons term, $g(\chi)F\wedge F$. 
When the scalar field $\chi$ is identified with 
the ghost or the quintessence field, this term affects the CMB polarizations  
after photon decoupling, through the rotation of the photon's 
polarization axis. As a result, we obtain non-vanishing TB- and EB-modes like  
$C^{\rm TB}_{\ell}=C^{\rm TE}_{\ell}\sin2\alpha$, where $\alpha$ is the rotation 
angle of the polarization axis \cite{Lue:1998mq,Feng:2006dp,Liu:2006uh,Cabella:2007br}. 
This is even true in the absence of tensor fluctuations, since a 
non-vanishing contribution is still obtained from scalar type fluctuations. 
Thus, for a small tensor-to-scalar ratio, the shape of TB-mode power spectrum is 
essentially the same as that of the scalar-type TE-mode spectrum. In this respect, 
a non-vanishing TB-mode spectrum by the electromagnetic Chern-Simons term 
may be clearly distinguished from that produced from circularly polarized GWBs. 
Note that the non-vanishing TB-mode is also obtained by the Faraday rotation 
through 
intervening magnetic fields \cite{Scannapieco:1997mt}. The Faraday rotation depends 
on the CMB photon frequency \cite{Kosowsky:1996yc} and it also alters the 
angular dependence of the TB-mode power spectrum. In this paper, we do 
not consider these two effects and just focus on the CMB power spectra 
from the circular polarized GWB.

\section{TB- and EB-mode power spectra from a circularly polarized GWB} 
\label{sec4}

We now consider the amplitude and the shape of the TB- and EB-mode power spectra 
discussed in \S 3, taking into account the secondary anisotropies. 
We will show that the effect of reionization 
greatly enhances the amplitude of the TB-mode at large angular scales. On the 
other hand, the effect of weak lensing is shown to be negligibly small. In 
this and following sections, we adopt 
the following cosmological parameters as fiducial model parameters, which are 
taken from the best-fit values of the three year WMAP data ($\Lambda$CDM+tensor), 
except for the tensor-to-scalar ratio $r=0.1$:
\ba
 \Omega_{\rm b}h^{2}=0.0233,\ \Omega_{\rm CDM}h^{2}=0.0962,\ 
\Omega_{\rm K}=0,\ h=0.787,\ 
\nonumber\\
 \tau_{\rm ri}=0.09,\ 
\Delta_{\mathcal{R}}^{2}(0.002/\mathrm{\rm Mpc})=2.1\times 10^{-9}\ ,
n_{\rm S}=0.984,\ r=0.1.
\ea
For simplicity, we assume the slow-roll consistency relation, $n_{\rm T}=-r/8$, 
and the vanishing running spectral index. The power spectra of CMB anisotropies 
presented here are all calculated based on the CAMB code \cite{Lewis:1999bs}, 
with suitable modification to compute TB- and EB-mode spectra.

 \subsection{Primary anisotropies}
 \label{sec:general_feature}

Let us first show the primary anisotropies of the TB- and EB-mode power 
spectra originating from the circularly polarized GWB of a primordial origin. 
\par

\begin{figure}
\begin{center}
\includegraphics[width=0.7\textwidth]{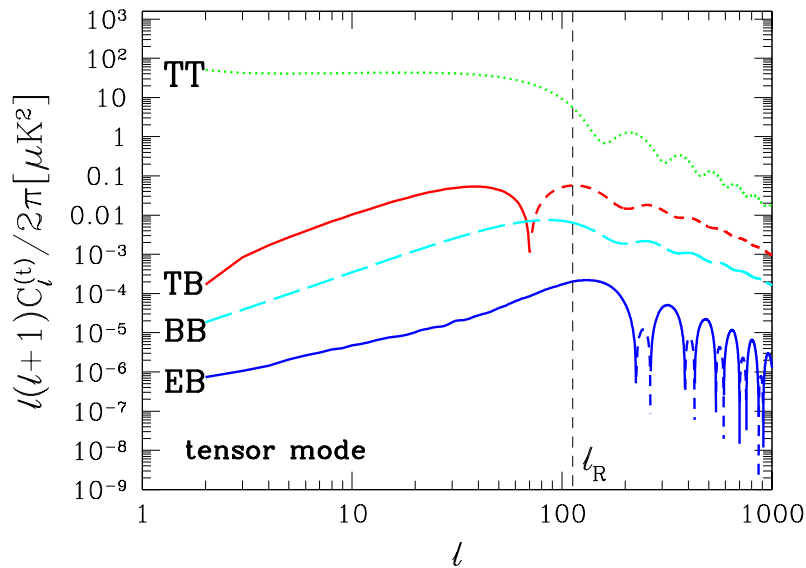}
\end{center}
\caption{The temperature and polarization cross spectra, 
$C_{\ell}^{{\rm TB}(t)}$ and $C_{\ell}^{{\rm EB}(t)}$, from circularly 
polarized gravitational waves. Here, setting the 
reionization optical depth, $\tau_{\rm ri}=0$, the absolute values of the cross power 
spectra are plotted for fiducial model with $\varepsilon=+1$.
In these plots, the negative correlation is indicated by 
the short-dashed lines.  As a reference, TT ({\it dotted}) and 
BB-mode ({\it long-dashed}) power spectra are also plotted. 
The vertical line labeled by $\ell_R (\sim 100)$, 
roughly corresponds to the angular size of the horizon radius 
at recombination epoch.}
\label{fig.1}
\end{figure}

In Figure \ref{fig.1}, specifically setting the parameter $\varepsilon=1$ 
corresponding to the right-handed polarized GWB, 
the TB- and EB-mode power spectra are plotted under 
the fiducial cosmological model except for the re-ionization parameter, 
$\tau_{\rm ri}=0$.  The results are then compared with the TT- and BB-mode spectra 
for the tensor fluctuations\footnote{Note that the sign of the TB-mode power 
spectrum plotted here is opposite to the one in Ref.\cite{Lue:1998mq}. Perhaps, 
this differences come from the definition of polarization bases, $e_{ab}^{\rm R,L}$. 
Our definition follows that of  Ref.\cite{Alexander:2004wk},  
i.e., equations (\ref{eq:def_R-handed}) and (\ref{eq:def_L-handed}). }. 
\par

Similarly to the BB-mode power spectrum, 
the TB- and EB-mode spectra have a peak at $\ell\sim \ell_{R}$, corresponding 
to the horizon scale at recombination (for details of the location of the BB-mode 
peak, see \cite{Pritchard05}). Also, at higher multipoles with $\ell> 200$,   
oscillatory behavior appears, which simply   
reflects the oscillations of the gravitational waves after the horizon 
re-entry time during the recombination epoch. A closer look at cross spectra 
reveals that while the EB-mode spectrum has many crossing points at higher 
multipoles, the TB-mode spectrum has one crossing 
point and the sign of the spectra is only changed around $\ell\sim 70$.   
Further, the amplitude of the EB-mode spectrum is extremely 
small compared to the one naively expected from the BB-(EE-)mode tensor spectrum. 
These characteristic behaviors basically come from the projection 
factors in the photon's transfer function $\Delta_{{\rm X}\ell}^{t}(k)$ (X=T, E, B). 
In \ref{sec:TBEB}, the reasons for these properties are discussed 
in some detail.
\par

\begin{figure}
\begin{center}
\includegraphics[width=0.49\textwidth]{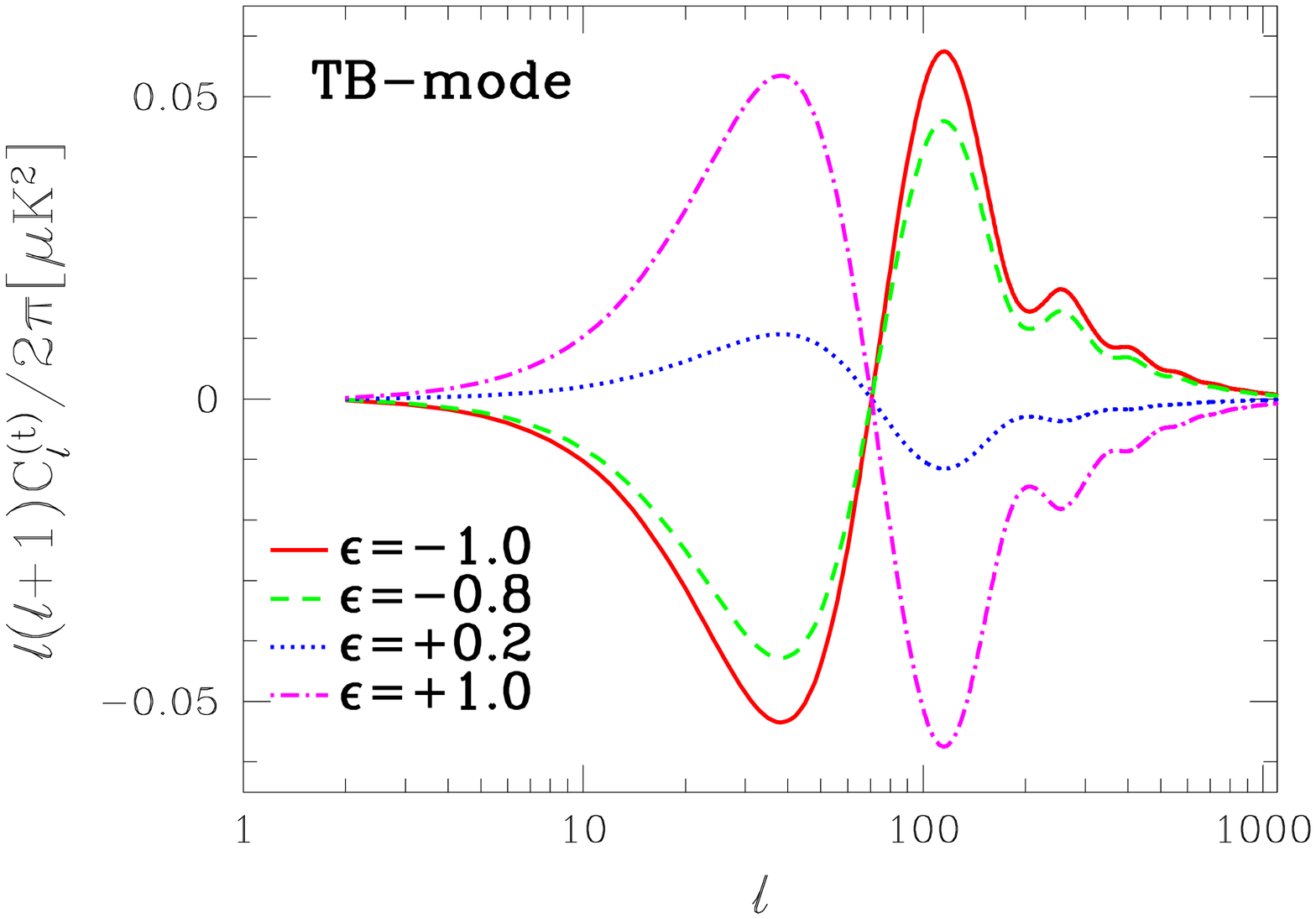}
\includegraphics[width=0.49\textwidth]{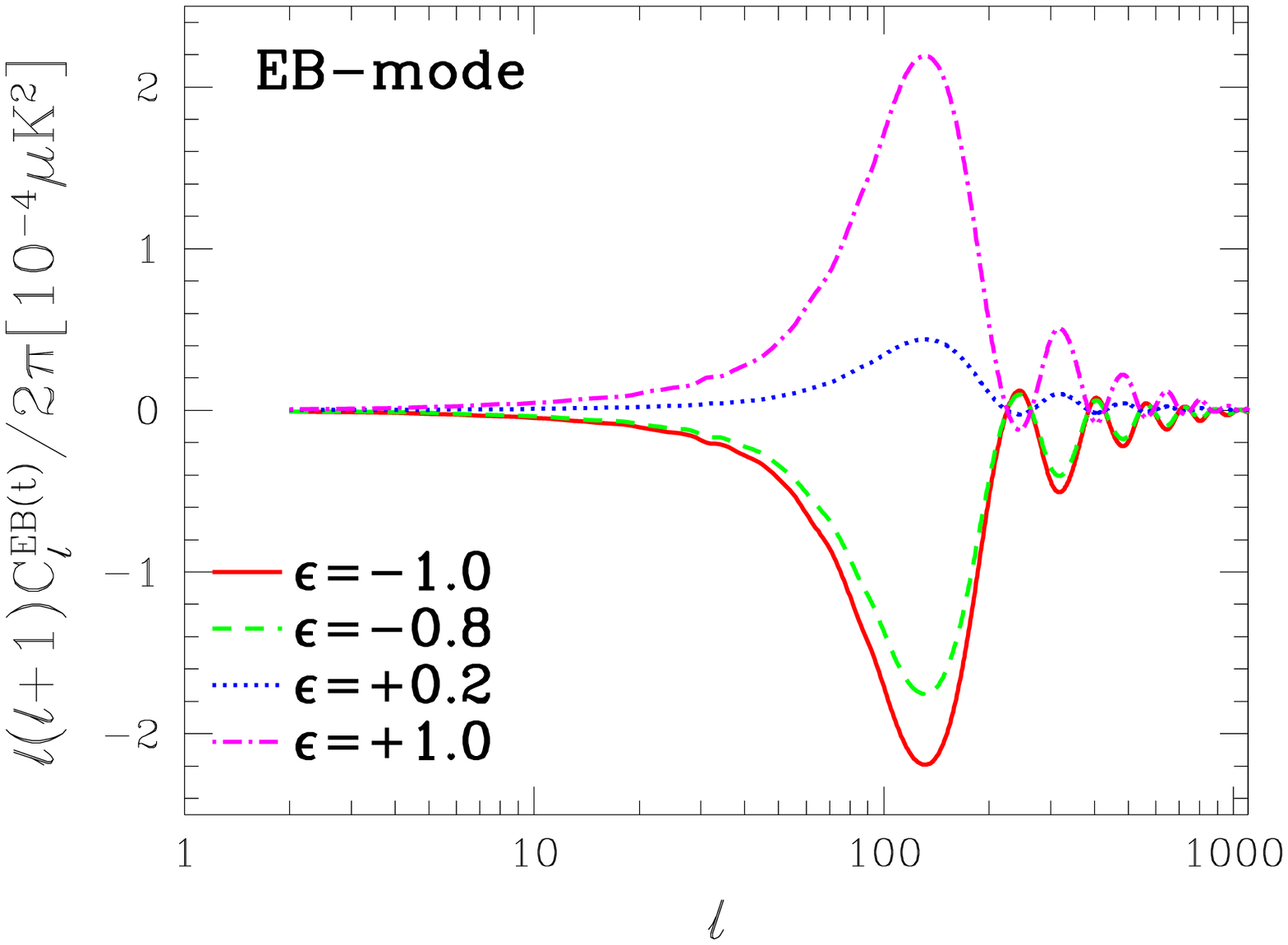}
\end{center}
\caption{Dependence of circular polarization degree, $\varepsilon$, on 
the TB- ({\it left}) and EB-mode ({\it right}) power spectra  
for the fiducial model except for the reionization optical depth, 
$\tau_{\rm ri}=0$.}
\label{fig.2}
\end{figure}

\begin{figure}
\begin{center}
\includegraphics[width=0.49\textwidth]{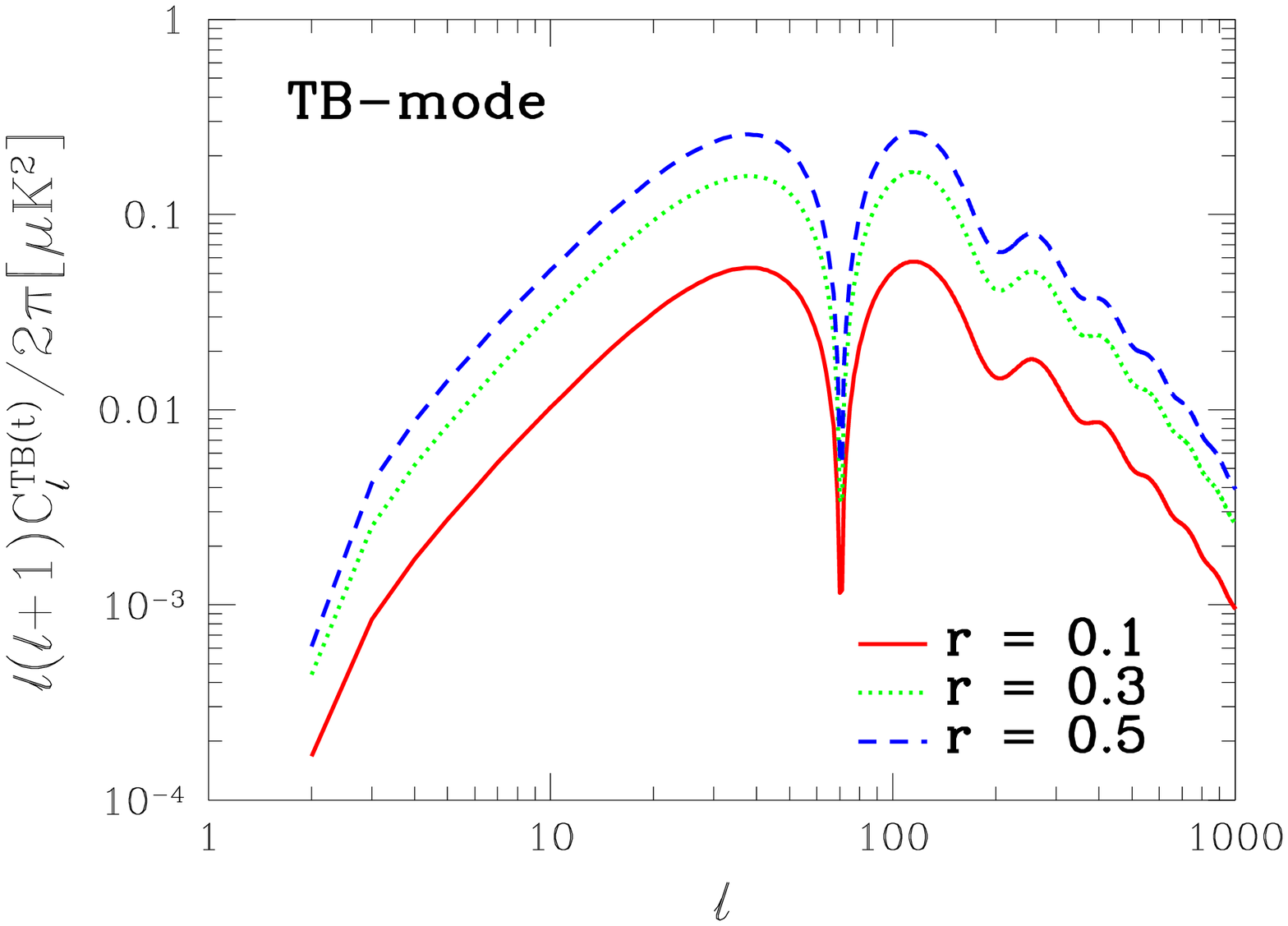}
\includegraphics[width=0.49\textwidth]{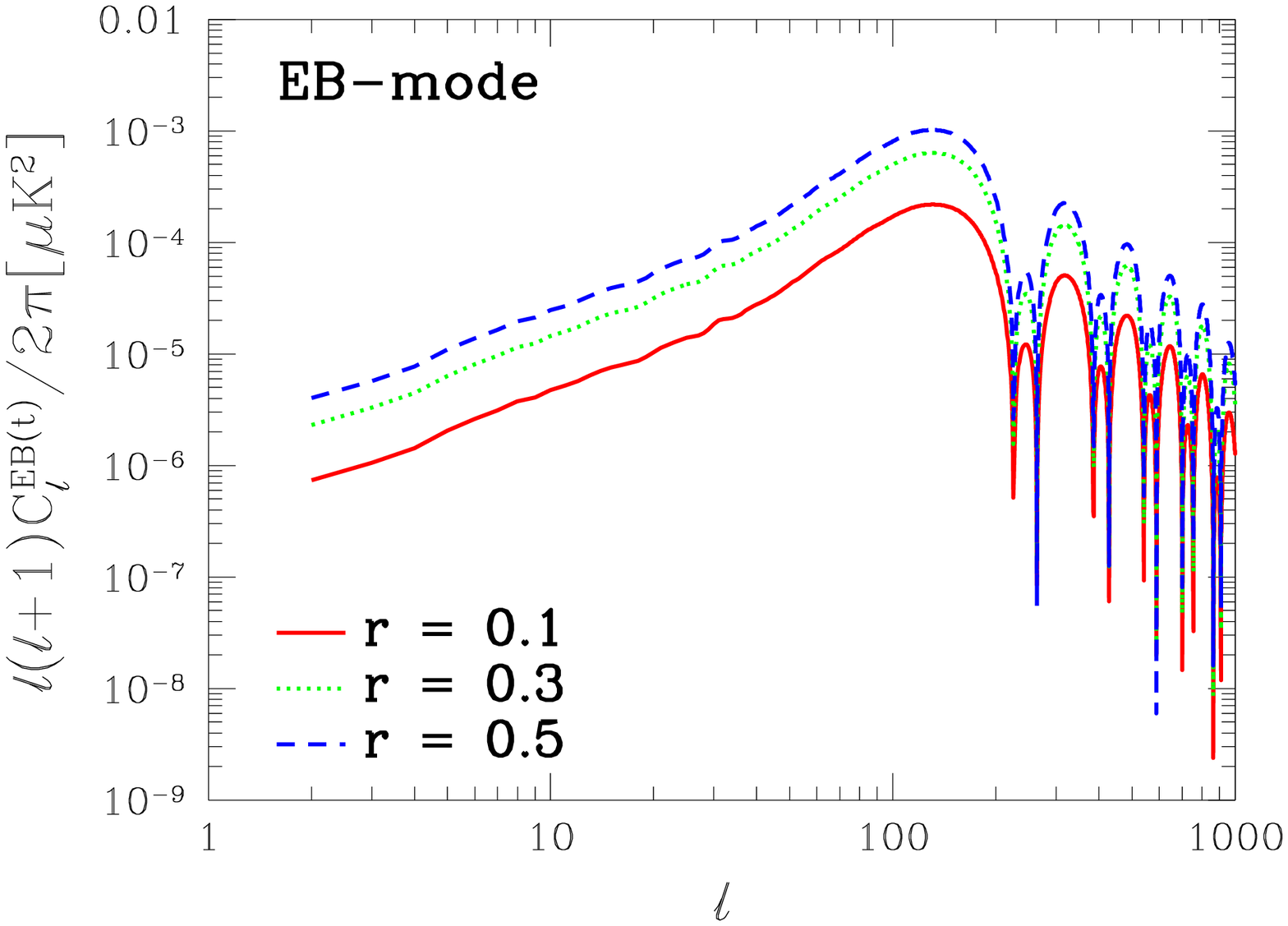}
\end{center}
\caption{Dependence of the tensor-to-scalar ratio, $r$, on 
 the TB- ({\it left}) and EB-mode ({\it right}) power spectra  
for the fiducial model with $\varepsilon=-1.0$. In these plots, 
the reionization optical depth is set to $\tau_{\rm ri}=0$, keeping 
the slow-roll consistency relation $n_{\rm T}=-r/8$.}
\label{fig.3}
\end{figure}

Apart from a tiny dependence on the density parameters such as 
$\Omega_{\rm b}$ and $\Omega_{\rm \Lambda}$ (e.g., 
see Ref.\cite{Zhang:2005nv} in the case of the BB-mode spectrum), 
the amplitude of primary TB- and EB-mode spectra are mainly determined 
by the tensor-to-scalar ratio $r$ and the fractional power of circular 
polarization $\varepsilon$. In Figures \ref{fig.2} and \ref{fig.3}, the dependence 
of the TB- and BE-mode spectra on $\varepsilon$ and $r$ (or $n_{\rm T}$) are 
shown respectively. Both of the parameters $\varepsilon$ and $r$ linearly 
alter the amplitude of spectra, 
but the degree of circular polarization, $\varepsilon$, allows us to change 
the over-all sign. 
This is the key to discriminate the polarization states of the GWB. 
Note that in plotting Figure \ref{fig.3},  
we strictly keep the slow-roll consistency relation, $n_{T}=-r/8$. However,  
the change of the spectral shape is very small and it would be difficult to   
observe it. 
\par
 

 \subsection{Effects of secondary anisotropies}

Let us move to the discussion on the effects of secondary anisotropies  
generated after the recombination epoch. 
\par

There are two possible sources to produce a large-angular scale anisotropy: 
reionization and the weak lensing. 
Among these, the weak lensing effect represents the gravitational deflection 
of a photon's propagation direction by the large-scale structure and it distorts 
the temperature and polarization maps of the CMB (see \cite{Lewis:2006fu} for a review). 
In particular, the effects from weak lensing are known as the big obstacle to detect the 
gravitational waves from the BB-mode power spectrum, since weak lensing newly 
creates the  B-mode polarization anisotropy from the scalar-type perturbations, 
which would dominate over the tensor fluctuation at $\ell\gtsim 100$. 
In the case of temperature-polarization cross spectra, however, 
transformation properties of E- and B-modes do not allow the production of 
a new TB-mode correlation from the scalar-type perturbations. This is also 
true for the EB-mode spectrum. As a result, the lensing effects on the TB- and 
EB-mode spectra are negligibly small. Detailed discussion on the effects of weak lensing 
are presented in \ref{sec:weak_lensing}.  
\par

\begin{figure}
\begin{center}
\includegraphics[width=0.495\textwidth]{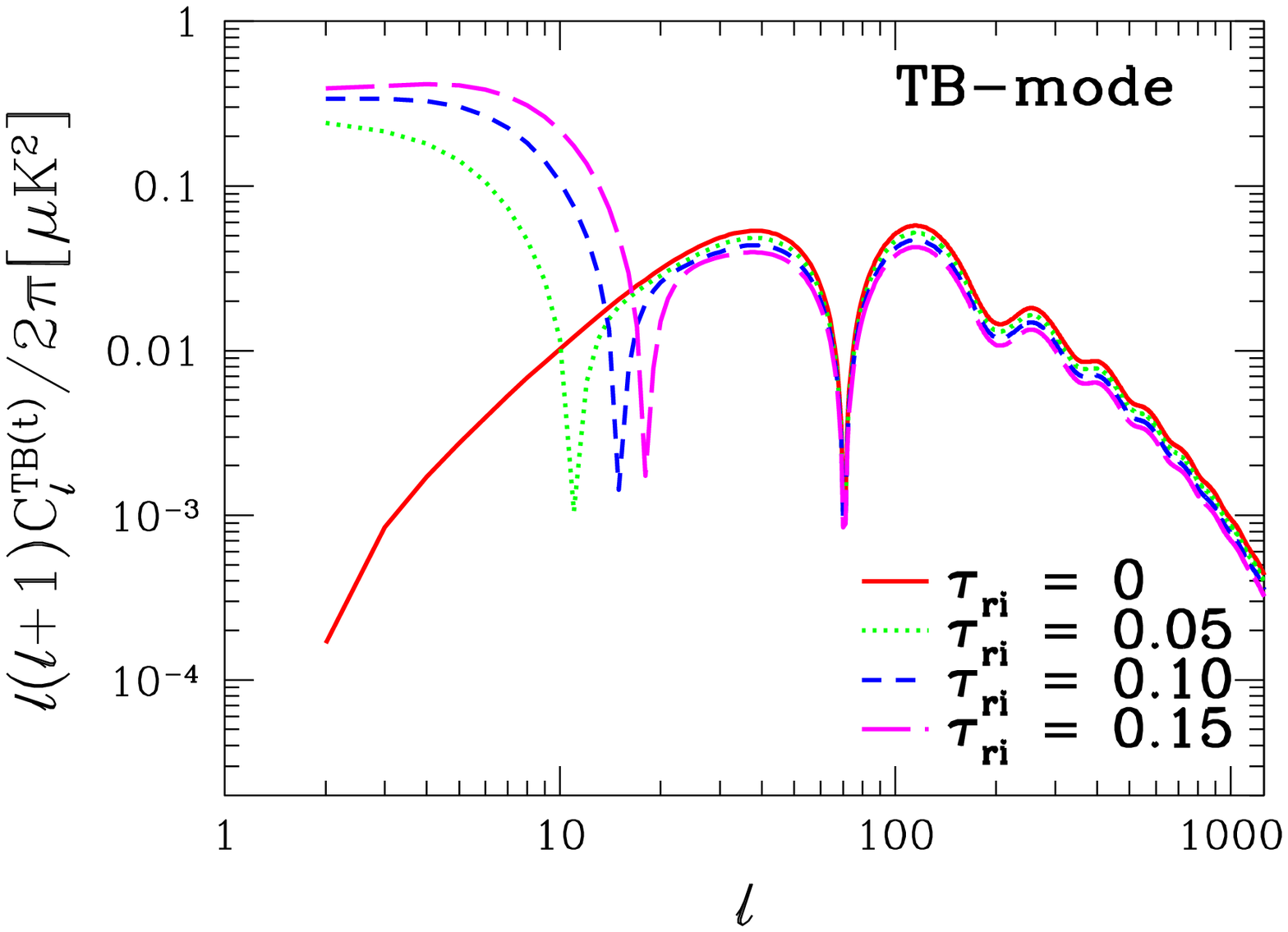}
\includegraphics[width=0.495\textwidth]{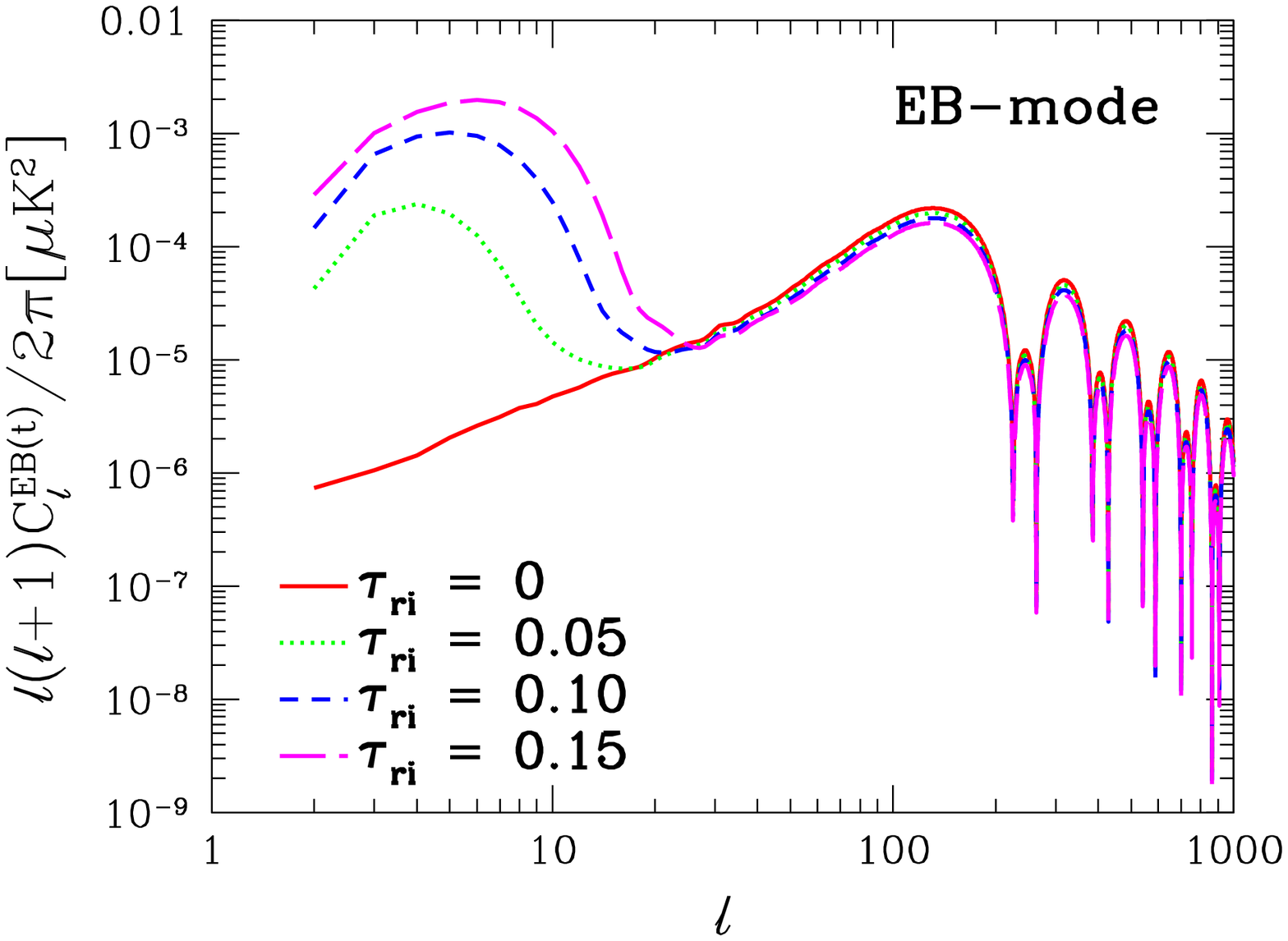}
\end{center}
\caption{Dependence of the reionization optical depth, $\tau_{\rm{ri}}$
on the TB- ({\it left}) and EB-mode ({\it right}) power spectra 
for the fiducial model with  $\varepsilon=-1.0$. 
A large enhancement of the amplitude arises due to the 
re-scattering of the CMB photons during reionization. 
}
\label{fig.4}
\end{figure}

On the other hand, the reionization of the universe drastically changes the 
large-scale anisotropies. Although 
details of the reionization history are model-dependent and are currently 
uncertain, its effect on the CMB power spectra is mainly characterized by the 
optical depth to the beginning of reionization, $\tau_{\rm ri}$ 
\cite{Basko80,Zaldarriaga:1996ke}. 
In Figure \ref{fig.4}, we show the TB- and EB-mode power spectra for various 
values of the reionization optical depth. Similar to the polarization 
spectra of scalar-type perturbations, the resultant power spectra are 
greatly amplified and a larger value of $\tau_{\rm ri}$ leads  
to a large amplitude of TB- and BE-mode at lower multipoles. This is essentially 
the same reason as in the scalar TE- and EE-mode spectra 
that the polarization anisotropies of the CMB photon are 
newly created from a primary anisotropy by Thomson scattering at 
reionization. The power spectra are sharply peaked at 
large-angular scales and the peak position $\ell_{\rm ri}$ 
is roughly estimated as $\ell_{\rm ri}\sim \sqrt{z_{\rm ri}}$  
\cite{Zaldarriaga:1996ke}. One interesting observation is that 
a new zero-crossing point appears around $\ell\sim10-20$ in the TB-mode 
spectrum and the amplitude of lower multipoles $\ell<6$ eventually 
changes its sign. 
\par

Although the precise form of the spectrum is difficult to predict analytically, 
the peak height of the spectrum caused by the reionization is roughly 
estimated as follows. First of all, the reionization reduces 
the fraction of photons reaching us from the recombination epoch. This 
is basically proportional to $\exp(-\tau_{\rm ri})$. Further, 
in the simple approximation with instantaneous reionization, 
the visibility function $g(\eta)=\tau'\,e^{-\tau}$ in equation 
(\ref{eq:visibility}) has a sharp peak around the reionization redshift 
$z_{\rm ri}$, in addition to the primary peak around the recombination 
epoch. These effects explicitly appear in 
the photon transfer function $\Delta^t_{{\rm X}\ell}$ or 
$\widetilde{\Delta}^t_{\rm X}$. Keeping this point in mind, 
from equation (\ref{eq:A8}), the transfer function for temperature 
fluctuation becomes
 \be
 \widetilde{\Delta}^{t}_{\rm T}(k,\mu)\simeq-\int^{\eta_{0}}_{0}
d\eta \,e^{i\,\mu\,k\,(\eta-\eta_0)}\,h'e^{-\tau}
\approx e^{-\tau_{\rm ri}}\widetilde{\Delta}^{t{\rm NR}}_{\rm T}\ ,
 \ee
where we have only considered the dominant term. Here, 
$\widetilde{\Delta}^{t{\rm NR}}$ represents the transfer function in the absence of 
reionization. In a similar manner, the transfer function for polarization fluctuations 
$\widetilde{\Delta}^t_{\rm P}$, given by (\ref{eq:A11}), 
is approximately described by 
\be
 \widetilde{\Delta}^{t}_{\rm P}(\eta_{0},k,\mu)=\int^{\eta_{0}}_{0}d\eta\, 
e^{i\,\mu\,k\,(\eta-\eta_0)}(-g\,\Psi) \approx \frac{1}{10}[1-e^{-\tau_{\rm ri}}]\,
\widetilde{\Delta}^{t{\rm NR}}_{\rm T}\ .  
\ee
Here, the source function $\Psi$ has been roughly evaluated from the monopole 
component of the temperature fluctuation as 
$\Psi\simeq(1/10)\,\widetilde{\Delta}_{{\rm T}0}^t$.  
The prefactor, $[1-e^{-\tau_{ri}}]$, indicates the fractional probability 
of photons scattered after the reionization before reaching the observer, 
leading to a new polarization anisotropy. 
\par

Based on the above approximations, the peak height of 
the power spectra is roughly estimated around $\ell\sim2$.  
From equations (\ref{eq:A1})--(\ref{eq:A7}), we obtain  
\ba
 C^{{\rm TT}(t)}_{\ell\sim 2} &\approx& e^{-2\tau_{\rm ri}}\,
C^{{\rm TT}(t){\rm NR}}_{\ell\sim 2}\ ,
\label{eq:approx_TT}
\\
 C^{{\rm EE}(t)}_{\ell\sim 2} &\approx& 
\frac{1}{100}[1-e^{-\tau_{\rm ri}}]^{2}\,C^{{\rm TT}(t){\rm NR}}_{\ell\sim 2}\ ,
\label{eq:approx_EE}
\\
 C^{{\rm BB}(t)}_{\ell\sim 2} &\approx& 
\frac{1}{100}[1-e^{-\tau_{\rm ri}}]^{2}\,C^{{\rm TT}(t){\rm NR}}_{\ell\sim 2}\ ,
\label{eq:approx_BB}
\\
 \left| C^{{\rm TB}(t)}_{\ell\sim 2} \right| &\approx& 
\frac{|\varepsilon|}{10} e^{-\tau_{\rm ri}}[1-e^{-\tau_{\rm ri}}]\,
C^{{\rm TT}(t){\rm NR}}_{\ell\sim 2}\ ,
\label{eq:approx_TB}
\ea
 where $C^{{\rm TT}(t){\rm NR}}_{\ell}$ stands for the temperature power 
spectrum for tensor mode without reionization. For fiducial cosmological 
parameters, the amplitude of the TB-mode at $\ell\sim 2$ is evaluated as 
\be
 \left|C^{{\rm TB}(t)}_{\ell\sim 2}\right| \approx 4\,\times 10^{-1}\,\,
|\varepsilon| \left(\frac{r}{0.1}\right)\ [\mathrm{\mu K^{2}}]\ .
\ee
With an appropriate range of the reionization optical depth 
$0.05\ltsim\tau_{\rm ri}\ltsim0.15$, 
the approximations (\ref{eq:approx_TT})--(\ref{eq:approx_TB}) agree reasonably 
with numerical results of the power spectra.

\begin{figure}
\begin{center}
\includegraphics[width=0.94\textwidth]{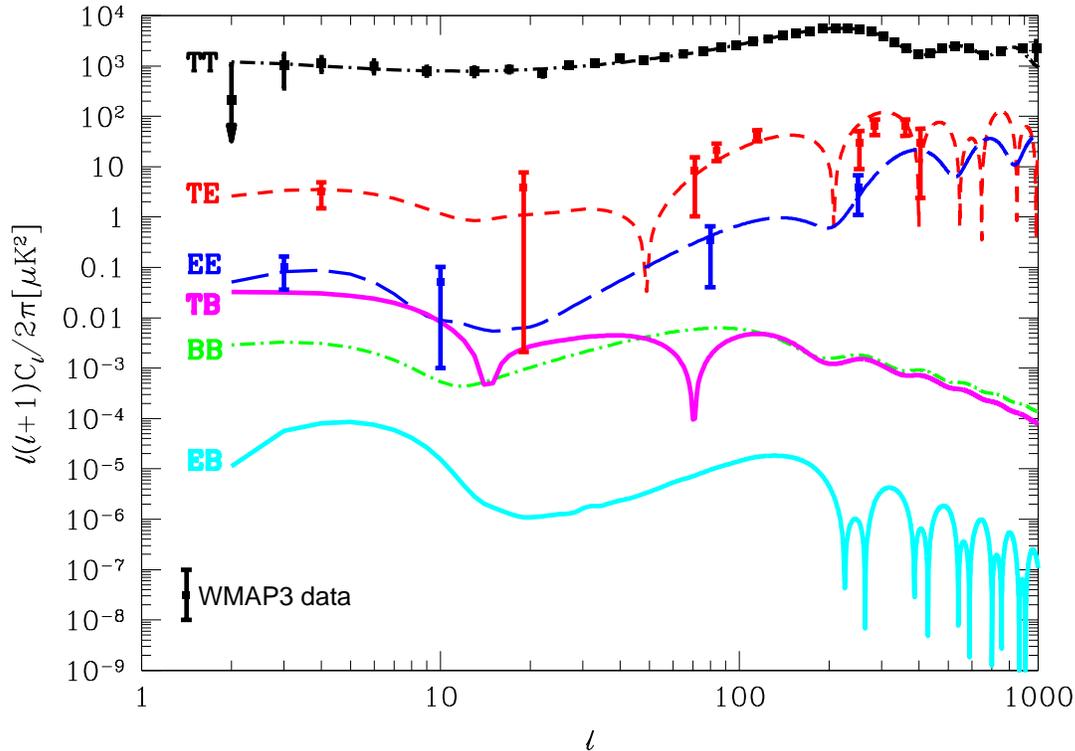}
\end{center}
\caption{CMB power spectra for the fiducial cosmology with $\varepsilon=+0.1$, 
including both scalar and tensor mode.
For comparison, three year WMAP data of TT, TE and EE-mode power spectra are plotted.
 At large-angular scale, the amplitude of 
the TB-mode ({\it magenta, solid}) exceeds 
 the BB-mode power spectrum ({\it green, dot-short dashed}) 
  and becomes comparable to the EE-mode power 
spectrum({\it blue, long-dashed}).}
\label{fig.5}
\end{figure}

\bigskip
As a summary of this section, we present the full CMB power spectra, i.e., 
the combined 
results of the contributions from both the scalar- and tensor-type perturbations. 
Figure \ref{fig.5} shows the results, specifically choosing the degree of polarization
as $\varepsilon=0.1$. With a slightly large value of the tensor-to-scalar ratio $r=0.1$, 
the amplitude of the TB-mode spectrum becomes comparable to the EE-mode spectrum. 
Interestingly, the amplitude of the TB-mode spectrum also exceeds the BB-mode 
amplitude at large angular scales. This is even true for small degree of polarization, 
$\varepsilon$\gtsim 0.01.

\section{Observational constraints on the circular polarization of the GWB}
\label{sec5}

Having understood the basic properties of the TB- and EB-mode power spectra,  
we now proceed to discuss the observational aspects for  
detecting a circularly polarized GWB. In section \ref{subsec:constraint_WMAP},  
for illustrative purposes, constraint on the degree of polarization of the 
GWB is derived 
based on the three year WMAP data.  In practice, we must wait for 
future polarization experiments in order to get a meaningful constraint. 
In section \ref{sec:Future prospects}, future prospects for measuring the 
degree of polarization of the GWB are addressed based on a Monte Carlo analysis of 
parameter estimation.

 \subsection{Constraints from three year WMAP data}
\label{subsec:constraint_WMAP}

Previous sections reveal that the effect of reionization largely amplifies 
the lower-multipole anisotropies and the amplitude of the TB-mode spectrum 
at multipoles $\ell\sim 2-10$ would be a clear indicator to measure the 
polarization states of the GWB. While currently no definite detection of 
the tensor-type fluctuations has been reported,  it is a good exercise 
to consider how one can constrain the circularly polarized GWB from 
lower-multipole data. For this purpose, we use the TB- and EB-mode power 
spectra taken from the three year WMAP data, currently the highest precision 
data at large-angular scales \cite{Page:2006hz}. Here, 
particularly using the lower-multipole data 
of $\ell\leq16$,  we perform a global parameter estimation together with 
the TT-, EE- and TE-mode data. We use the publicly available Markov-Chain Monte 
Carlo code, COSMOMC 
\cite{Lewis:2002ah}, which we modified to compute the TB- and EB-mode power 
spectra originating from a circularly polarized GWB. 
\par

In the present analysis, we use the likelihood function for 
TT-, EE- and TE-mode spectra available on the LAMBDA website 
\footnote{{\tt http://lambda.gsfc.nasa.gov/product/map/dr2/likelihood\_get.cfm}}. 
As for the TB- and EB-mode data, we simply assume the Gaussian likelihood 
function: 
\be
 \mathcal{L}_{\rm TB/EB} = \exp \left(-\frac{\chi^{2}_{\rm TB/EB}}{2}\right)\ ,
\label{eq:TBEBlikelihood}
\ee
with
\be
 \chi^{2}_{\rm TB/EB} = 
\sum_{\ell}\frac{(\overline{C}^{\rm TB/EB}_{\ell}-
\widehat{C}^{\rm TB/EB}_{\ell})^{2}}
{\sigma^{2}_{\ell}}\ ,
\ee
where the quantities $\overline{C}^{\rm TB/EB}$ and $\widehat{C}^{\rm TB/EB}$
respectively denote the theoretical value and the experimental data of the 
TB- or EB-mode power spectra. The quantity $\sigma_{\ell}^{2}$ denotes the 
variance of estimated power spectra at each multipole, corresponding to  
the diagonal component of the covariance matrix.  
Strictly speaking,  the assumption (\ref{eq:TBEBlikelihood}) 
is not valid for the three year WMAP data. 
For full-sky coverage, the exact likelihood function significantly deviates from 
the Gaussian likelihood function at lower multipoles \cite{Bond:1998qg}. 
Nevertheless, just 
for illustrative purpose, we adopt the Gaussian form (\ref{eq:TBEBlikelihood}), 
since we do not know the precise form of likelihood function suitable for WMAP 
experiment including TB- and EB-mode power spectra.
A more rigorous treatment including the non-Gaussianity in the likelihood 
function will be discussed in the next subsection. 
\par

To derive the constraint, we consider a spatially flat cosmology and treat 
the following eight parameters as free parameters: 
\be
 (\Omega_{\rm b}h^{2},\,\,\Omega_{\rm CDM}h^{2},\,\,\theta,\,\,\tau_{\rm ri},\,\,
n_{\rm S},\,\,A_{\rm S},\,\,r,\,\,\varepsilon)
\ee
where $\theta$ is the ratio of the sound-horizon scale to the angular 
diameter distance. The parameters $n_{\rm S}$ and $A_{\rm S}$ are 
the scalar spectral index and the amplitude of the curvature perturbation, 
respectively. Then, keeping the slow-roll consistency relation, 
$n_{T}=-r/8$, we perform a global estimation of the cosmological parameters.
\par

\begin{figure}
\begin{center}
\includegraphics[width=0.6\textwidth]{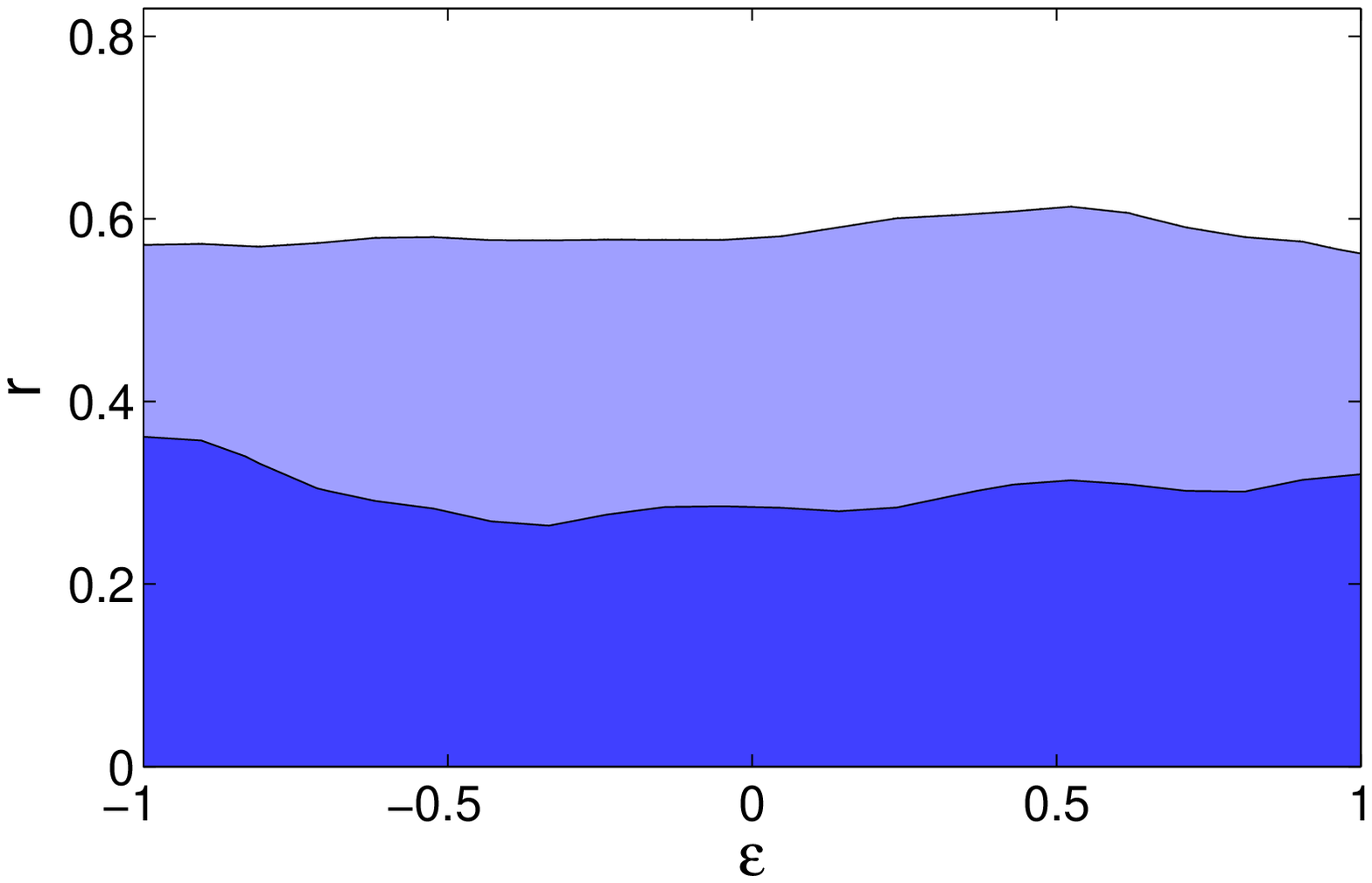}

\includegraphics[width=0.45\textwidth]{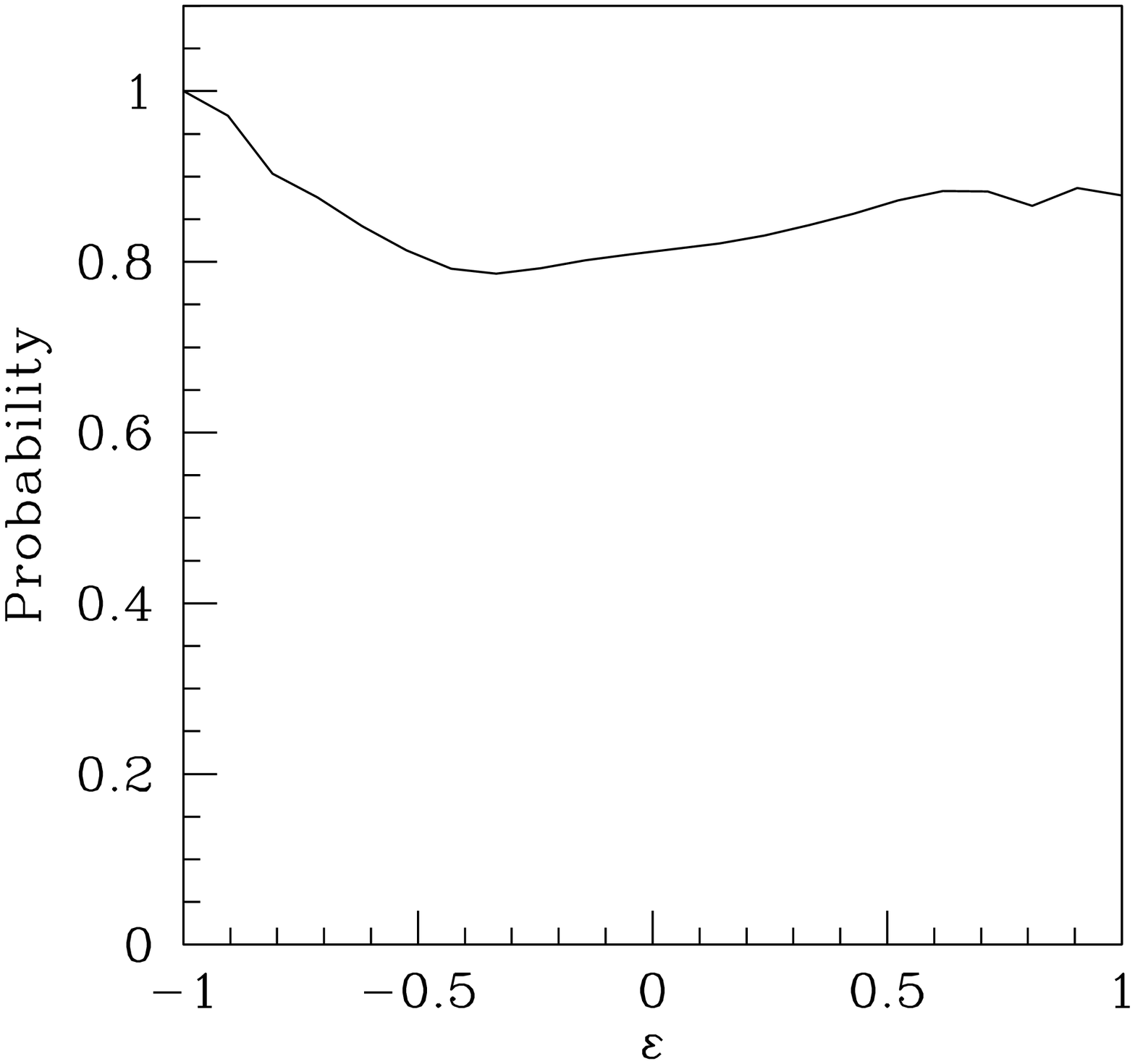}
\hspace*{0.5cm}
\includegraphics[width=0.45\textwidth]{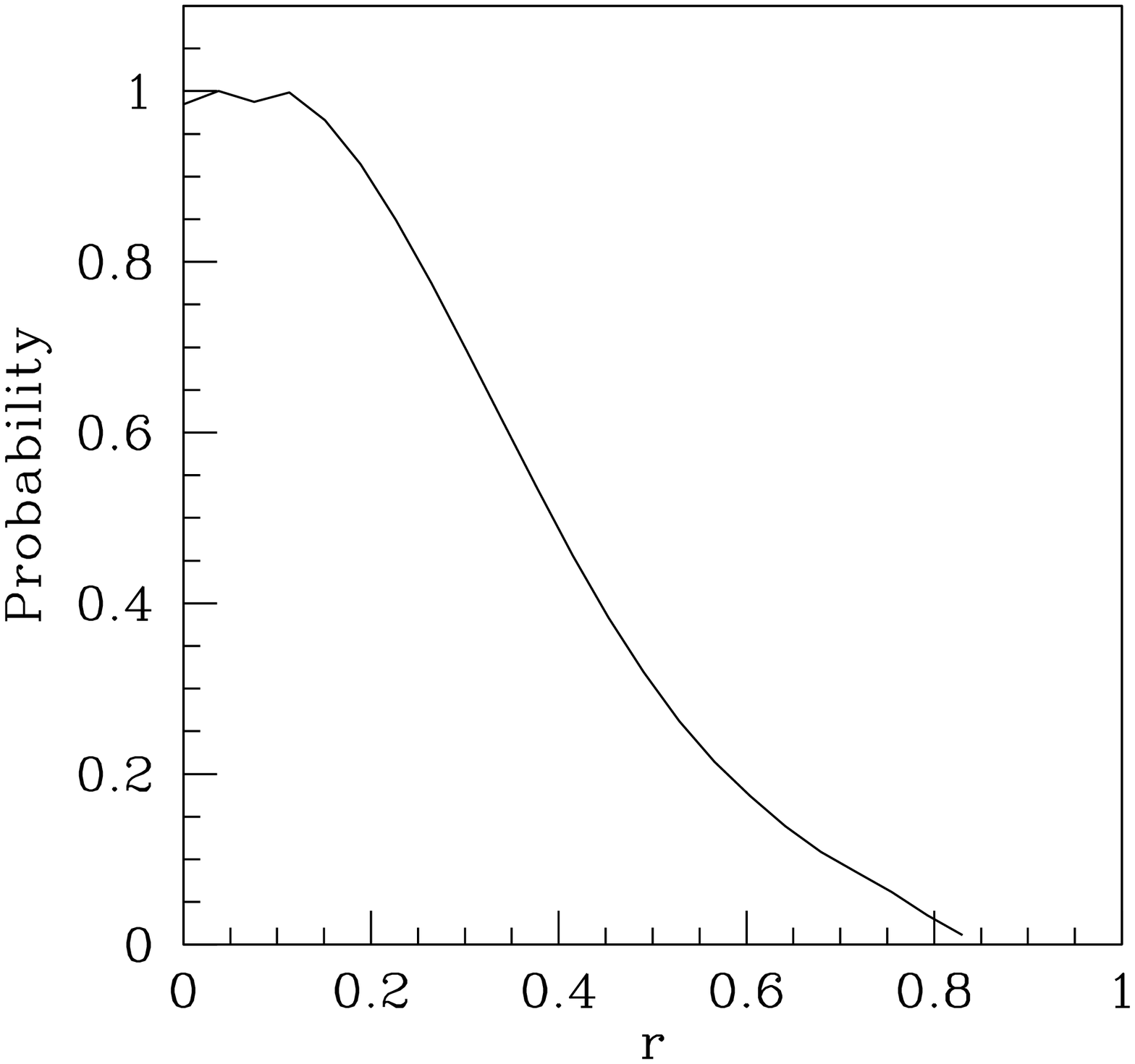}
\end{center}
\caption{Constraints on the circularly polarized GWB from the three year WMAP 
  data. Top panel shows the 68\% ({\it blue}) and the 95\% ({\it light-blue}) 
  confidence regions 
  of the parameters $r$ and $\varepsilon$. The results are obtained by marginalizing 
  over the other cosmological parameters. Bottom panel shows the posterior distribution 
  for the degree of polarization $\varepsilon$ ({\it left}) and the tensor-to-scalar 
  ratio $r$ ({\it right}). }
\label{fig.6}
\end{figure}

Figure \ref{fig.6} shows the constraints 
on the tensor-to-scalar ratio $r$ and the circular polarization degree 
$\varepsilon$ by marginalizing over the other cosmological parameters. 
Top panel plots the projected two-dimensional contours 
of 68\% (blue) and 95\% (light-blue) confidence regions, while bottom panels give 
the marginalized one-dimensional posterior distribution for  
the parameters $\varepsilon$ ({\it left}) and $r$ ({\it right}). 
Note that 
the constraints on the other cosmological parameters are also 
derived and our constraints reasonably match those obtained 
by the WMAP team. 
\par

From Figure \ref{fig.6}, no definite constraint on the degree of circular polarization 
 was obtained. This is simply because the uncertainty in 
the tensor-to-scalar ratio $r$ is still large. 
Although the 95\% confidence limit of 
$r$ is slightly reduced to $r<0.59$ compared to the WMAP 
result with $r<0.65$\footnote{This result is obtained using the three year WMAP 
data with TT-, TE- and EE-mode. Note that the tightest constraint is 
$r<0.30$ with WMAP3+SDSS.}, this is still consistent with the 
vanishing tensor-to-scalar ratio $r=0$. A closer look at the posterior 
distribution reveals that there is a local maximum around 
$\varepsilon\sim-1$. However, observational errors of the 
TB- and EB-mode spectra are very large and the agreement between 
theory and observation is not visually clear. 
Therefore, the significance of non-vanishing $\varepsilon$ is very low. 
We conclude that no meaningful constraint on the degree of circular polarization 
is obtained.

 \subsection{Future prospects}
 \label{sec:Future prospects}

Focusing on the prospects for measuring the circular 
polarization degree, we estimate the expected constraints derived 
from the future experiments. In what follows, assuming the complete 
subtraction of the foreground sources, we address principal aspects for 
detecting a circularly polarized GWB. 
We examine the two specific cases: 
forthcoming experiment by PLANCK satellite and a
cosmic-variance limited experiment idealistically corresponding to the 
next-generation CMB measurement. 
\par

As mentioned in the previous subsection, the Gaussian 
likelihood function for the $C_{\ell}$'s is an inadequate assumption 
at lower multipoles \cite{Bond:1998qg} and 
the non-Gaussianity arising from the 
cosmic variance should be properly taken into account. 
Further, notice the large cosmic variance 
for the TB-mode power spectrum. This 
is deduced from the diagonal component 
of the covariance matrix, ${\rm Cov}_{\ell}^{\rm TB}$: 
\be
 {\rm Cov}_{\ell}^{\rm TB} =
\frac{(\overline{C}^{\rm TT}_{\ell}+N^{\rm TT}_{\ell})
(\overline{C}^{\rm BB}_{\ell}+
N^{\rm BB}_{\ell})+(\overline{C}^{\rm TB}_{\ell})^{2}}
{(2\ell+1)f_{\mathrm{sky}}}\ ,
\label{eq:cov_diagonal_TB}
\ee
which roughly corresponds to the estimation error of the power spectrum. 
Here, $N^{\rm TT}_{\ell}$ and $N^{\rm BB}_{\ell}$ denote the experimental 
noises for temperature and polarization maps, and $f_{\mathrm{sky}}$ is 
the fractional  sky coverage. In the above expression, 
theoretical power spectrum $\overline{C}_{\ell}^{\rm TT}$ 
includes the contribution from both 
the scalar- and tensor-type perturbations. Thus, for a small 
tensor-to-scalar ratio, the dominant contribution to 
${\rm Cov}_{\ell}^{\rm TB}$ always comes from the first term 
$(\overline{C}^{\rm TT}_{\ell}+N^{\rm TT}_{\ell})
 (\overline{C}^{\rm BB}_{\ell}+N^{\rm BB}_{\ell})
 \simeq \overline{C}^{\rm TT}_{\ell}\overline{C}^{\rm BB}_{\ell}$, 
leading to a large uncertainty in the power spectrum estimation. 
This is true even in the absence of the primary TB-mode anisotropy. 
In this respect, definite detection 
of the degree of circular polarization requires a larger value of $\varepsilon$ and 
a proper treatment of the cosmic-variance is crucial to get the 
correct constraints.

Keeping the above remarks in mind,  we adopt the non-Gaussian likelihood 
function derived in \ref{sec:likelihood}: 
\ba
&& -2\ln \mathcal{L}  =  
\sum_{\ell}(2\ell+1)f_{\mathrm{sky}}
\left\{\ln\left(\frac{\overline{C}^{\rm TT}_{\ell}\overline{C}^{\rm BB}_{\ell}
-(\overline{C}^{\rm TB}_{\ell})^{2}}
{\widehat{C}^{\rm TT}_{\ell}\widehat{C}^{\rm BB}_{\ell}-
(\widehat{C}^{\rm TB}_{\ell})^{2}}\right)\right.
\nonumber\\
&&\quad\quad\quad\quad\quad\quad\quad\quad\quad\quad\left.
+\,\,\frac{\widehat{C}^{\rm TT}_{\ell}
\overline{C}^{\rm BB}_{\ell}+\overline{C}^{\rm TT}_{\ell}
\widehat{C}^{\rm BB}_{\ell}-2\overline{C}^{\rm TB}_{\ell}
\widehat{C}^{\rm TB}_{\ell}}
{\overline{C}^{\rm TT}_{\ell}\overline{C}^{\rm BB}_{\ell}-
(\overline{C}^{\rm TB}_{\ell})^{2}}-2\right\}\ .
\label{eq:exact_likelihood}
\ea
Again, the quantities $\overline{C}_{\ell}^{\rm XY}$ and 
$\widehat{C}_{\ell}^{\rm XY}$
respectively denote the theoretical and the estimated values of the power 
spectra. Note that the likelihood function (\ref{eq:exact_likelihood}) 
becomes maximum when 
$\overline{C}_{\ell}^{\rm XY}=\widehat{C}_{\ell}^{\rm XY}$. 
The above expression is the exact result for 
the experimental data with full-sky coverage $f_{\rm sky}=1$, but it 
still provides a good description for an experiment with almost full-sky 
coverage, like PLANCK. 
\par

Based on the likelihood function (\ref{eq:exact_likelihood}), 
we perform a likelihood analysis to estimate 
the sensitivity of future experiments for constraining the parameters 
$r$ and $\varepsilon$. To do this, we use the TT-, BB- and TB-mode  
power spectra with multipoles $\ell\leq 100$. 
The data points for each power spectrum $\widehat{C}_{\ell}^{\rm XY}$ 
are exactly set to the fiducial theoretical values. This is 
equivalent to the averaged data set over the infinite number of mock samples 
\cite{erotto:2006rj}. For the cosmic-variance limited experiment, 
we just use the form (\ref{eq:exact_likelihood}) and 
simply set $f_{\mathrm{sky}}$ to unity. On the other hand, 
for the PLANCK setup, both the theoretical and 
experimental power spectra in the likelihood function 
(\ref{eq:exact_likelihood}) are replaced with those including noise bias 
contributions as $\overline{C}^{\rm TT/BB}_{\ell}\to
(\overline{C}^{\rm TT/BB}_{\ell}+N^{\rm TT/BB}_{\ell})$ and 
$\widehat{C}^{\rm TT/BB}_{\ell}\to
(\widehat{C}^{\rm TT/BB}_{\ell}+N^{\rm TT/BB}_{\ell}$). 
The noise power spectra for the PLANCK experiment are given by
\be
 N_{\ell}^{\rm XX} = \omega^{-1}_{\rm X}W^{-2}_{\ell}
=(\sigma_{\rm P,X}\,\theta_{\rm FWHM})^{2}
\exp\left[\frac{\ell(\ell+1)}{\ell^{2}_{\mathrm{beam}}}\right]
\ee
with subscript $XX$ being $XX=TT\ \mathrm{or}\ BB$. The quantity 
$\omega_{X}$ is the weight factor per 
solid angle, $W_{\ell}$ is the beam window function, and the beam size, 
$\ell_{\mathrm{beam}}$, is given by $\ell_{\mathrm{beam}}=\sqrt{8\ln 2}/
(\theta_{\rm FWHM})$ for the Gaussian beam. 
For the average sensitivity per pixel, $\sigma_{P,X}$, and angular resolution, 
$\theta_{FWHM}$, we adopt the values for 
the high frequency instruments of 100, 143 and 217GHz 
channels (see Table 1.1 of \cite{Planck:2006uk} for instrumental performance). 
The sky coverage of PLANCK is assumed to be  
$f_{\mathrm{sky}}=0.65$, corresponding to a $\pm 20$ degrees Galactic cut. 
\par

\begin{figure}
\begin{center}
\includegraphics[width=0.55\textwidth]{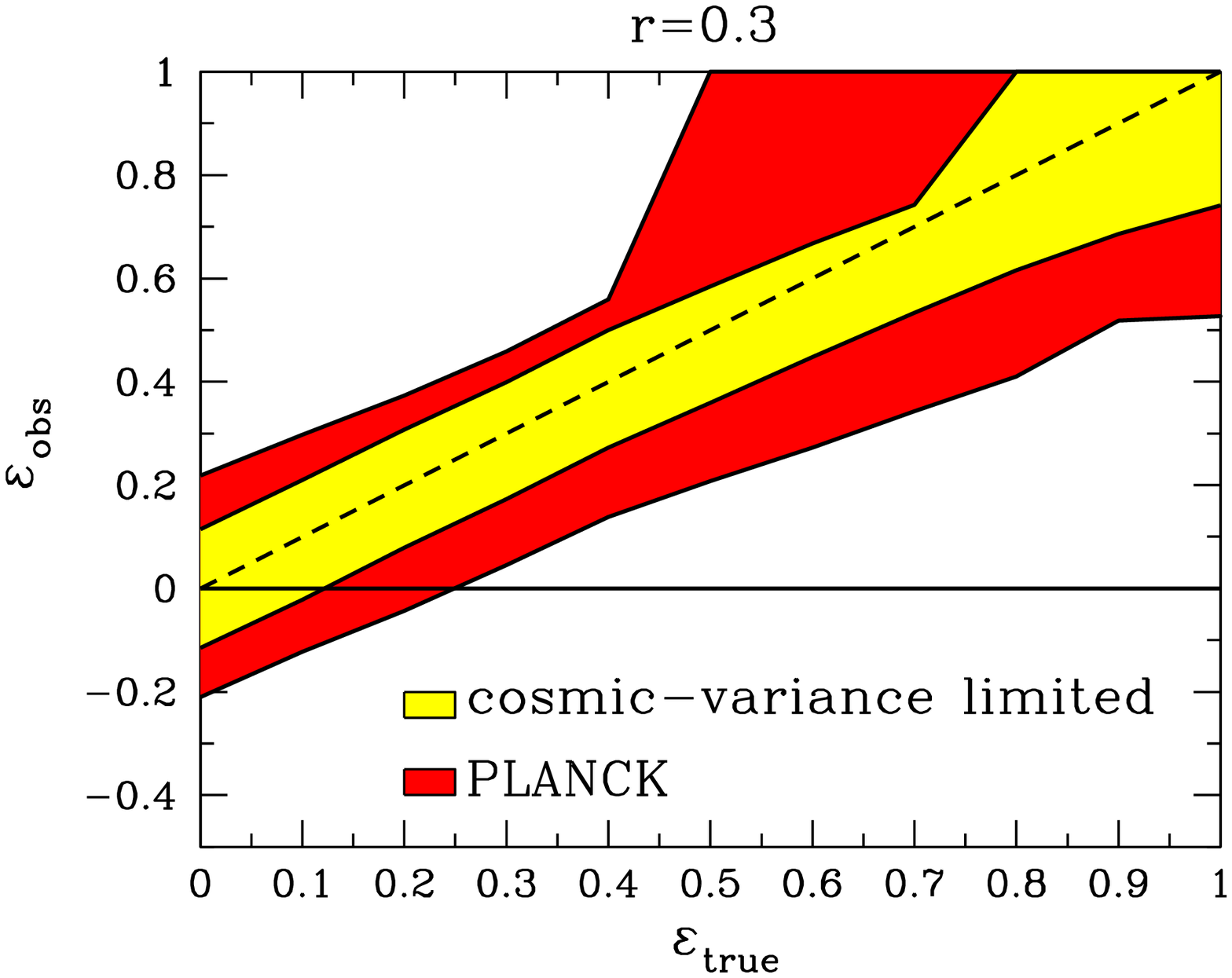}

\includegraphics[width=0.55\textwidth]{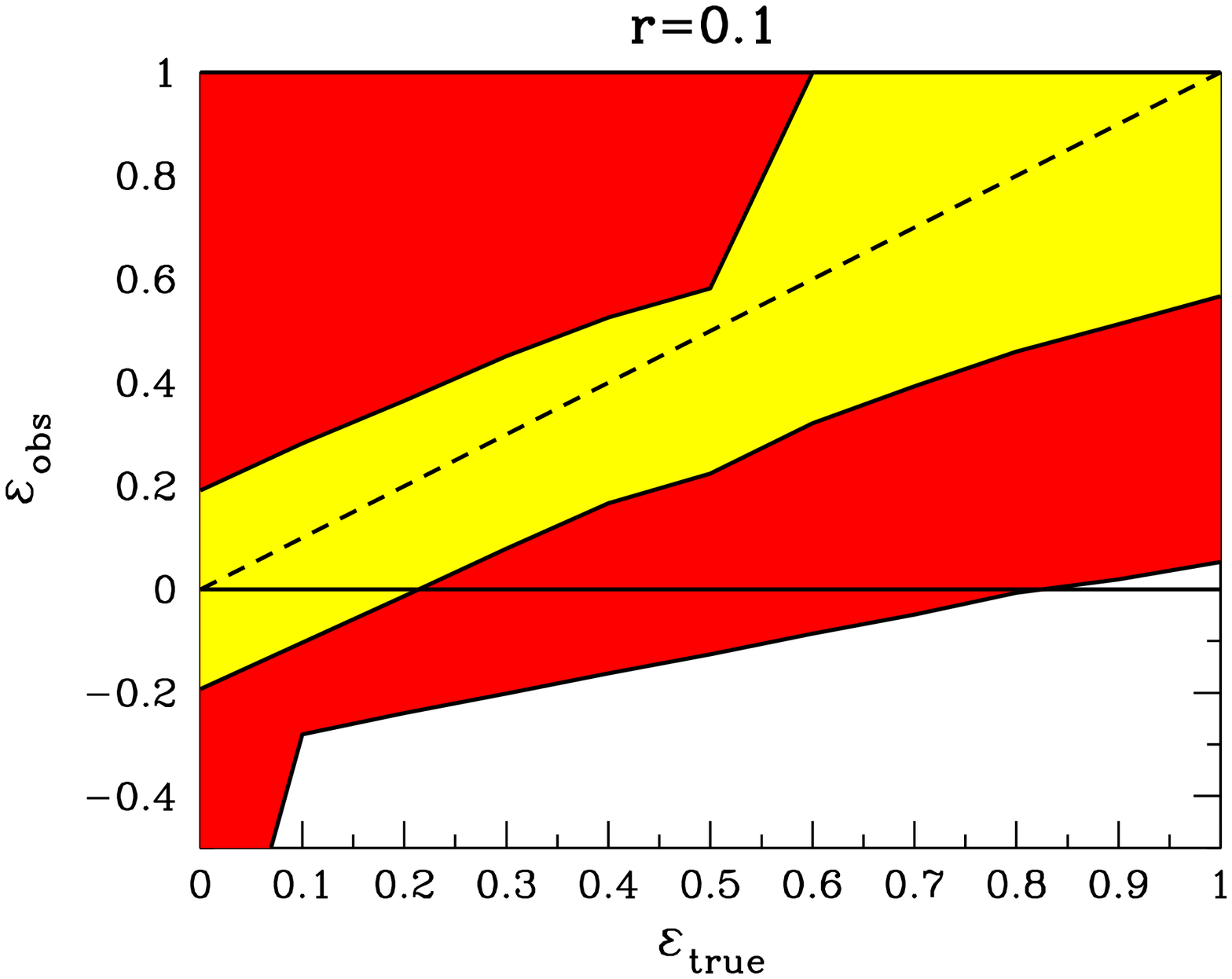}

\includegraphics[width=0.55\textwidth]{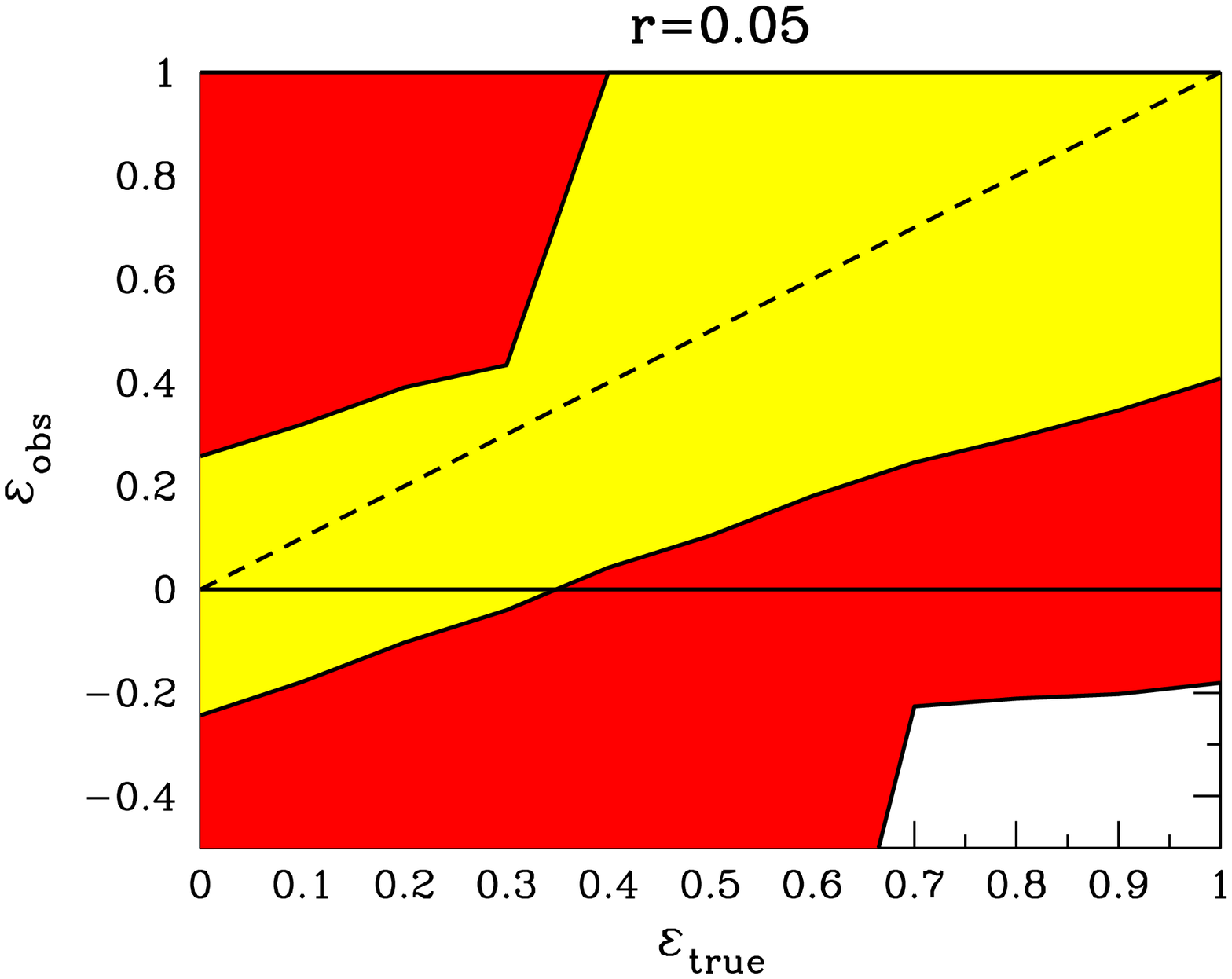}
\end{center}
\caption{Expected sensitivity of future experiment to the constraint 
on the circular polarization degree $\varepsilon$ for 
the specific tensor-to-scalar ratio: $r=0.3$({\it top}), $r=0.1$({\it middle}) 
and $r=0.05$({\it bottom}). In each panel, 
the one-dimensional marginalized 68\% confidence region of 
the estimated value of circular polarization degree, 
$\varepsilon_{\mathrm{obs}}$, is plotted as a function of the true 
input value, $\varepsilon_{\mathrm{true}}$. The {\it yellow} and 
{\it red} shaded region respectively represents the 
expected sensitivity of the PLANCK and ideal cosmic-variance 
limited experiments. The dashed line indicates 
$\varepsilon_{\mathrm{true}}=\varepsilon_{\mathrm{obs}}$. 
}
\label{fig.7}
\end{figure}

Figure \ref{fig.7} displays the results for the expected sensitivity of 
future experiments to the constraint on the degree of 
circular polarization $\varepsilon$ for 
the specific tensor-to-scalar tensor ratio: $r=0.3$ ({\it top}), $0.1$ 
({\it middle}) and $0.05$ ({\it bottom}).  In each panel, the marginalized 
68\% confidence regions of the posterior distribution for $\varepsilon$ 
are plotted for PLANCK ({\it red}) and cosmic-variance limited 
({\it yellow}) experiments, 
as a function of the true input value, $\varepsilon_{\rm true}$ 
\footnote{The upper and lower values of the 68\% confidence region,  
$[\varepsilon_1,\,\,\varepsilon_2]$, around the 
best-fit value are estimated from the marginalized posterior distribution 
$P(\varepsilon_{\rm obs})$ as 
$\int_{\varepsilon_1}^{\varepsilon_2}\,\,d\varepsilon_{\rm obs} 
 P(\varepsilon_{\rm obs}) =0.68$ under 
equi-probability, $P(\varepsilon_1)=P(\varepsilon_2)$.  
In cases with $\varepsilon_2$ ($\varepsilon_1$) reaching $1$ ($-1$),  
we simply set it to $1$ ($-1$).}. At first sight, a definite detection 
of the degree of circular polarization seems difficult for small 
tensor-to-scalar ratios. This is simply due to the large cosmic 
variance coming from the contribution 
$\overline{C}_{\ell}^{\rm TT}\overline{C}_{\ell}^{\rm BB}$ 
(see Eqs.(\ref{eq:cov_diagonal_TB}) and (\ref{eq:exact_likelihood})),  
in which the TT-mode spectrum $\overline{C}_{\ell}^{\rm TT}$ always 
gives a large value and is not much affected by the tensor-to-scalar 
ratio. From Figure \ref{fig.7}, the PLANCK experiment hardly constrains  
the degree of circular polarization at $r\ltsim0.1$, below which the 
68\% confidence level extends over the region $\epsilon_{\rm obs}<0$  
and one cannot clearly discriminate between 
polarized and un-polarized GWBs. 
\par

On the other hand, 
for the idealistic situation of cosmic-variance limited experiment, 
there still exists a window to distinguish a 
signature of circularly polarized GWB from the cosmic-variance 
dominated data. From Figure \ref{fig.7}, the detectable level
of the polarization degree can be read off:  
\be
\left|\varepsilon_{\rm obs}\right| \gtsim 
0.35 \left(\frac{r}{0.05}\right)^{-0.61}. 
\ee
Note that this estimate is roughly consistent with the one obtained by 
Ref.\cite{Lue:1998mq}, in which the authors reported that 
post-PLANCK experiment 
might conceivably be able to discriminate a value as small as 
$\varepsilon\sim0.08$ for the tensor-to-scalar ratio $r=0.7$,  
comparable to our estimate of the detectable level, 
$0.07$. However, they did not properly take into account the effects of 
reionization. Further, they only used the TB-mode spectrum 
to derive a minimum detectable $\varepsilon$. 
In this respect, close agreement between ours and Ref.\cite{Lue:1998mq} 
might be regarded as an accidental one. 
\par

Anyway, a realistic value of the tensor-to-scalar ratio is expected to be 
much smaller than unity. Our results imply that a large value of $\varepsilon$ 
is generally required in order to falsify the possibility of an un-polarized GWB. 
However, we do not theoretically exclude the possibility of a perfectly 
polarized GWB. Though difficult, it is still worthwhile to explore 
a signature of parity violation in the universe with future CMB experiments.

\section{Discussion and conclusions}
\label{sec6}

We have extensively discussed the detectability of the polarized states 
of primordial gravitational waves imprinted in the CMB anisotropies. 
In the early universe, 
the parity violation term originating from superstring theory or M-theory 
generically arises, which may produce a circularly polarized GWB during 
inflation. Such asymmetrically polarized gravitational waves 
induce a non-trivial correlation of CMB anisotropies between 
temperature and polarization modes. We have calculated the 
power spectra of CMB anisotropies generated from a circularly polarized 
GWB (i.e., TB- and EB-mode spectra). Taking into account the secondary 
anisotropies, we found that the effect of reionization creates a large 
amplitude of the lower multipoles of TB- and EB-mode spectra, 
which may be helpful to 
constrain the tensor-to-scalar amplitude ratio, $r$, as well as the degree of 
circular polarization of the GWB, $\varepsilon$. We then move to 
discuss observational aspects for detecting a circular polarized GWB. 
Using the three year WMAP data, we demonstrated how one can constrain the 
parameters $\varepsilon$ and $r$ from TB- and EB-mode data. 
For future prospects, we derive an expected sensitivity of representative 
experiments, i.e., PLANCK and cosmic-variance limited experiments, to the degree of 
the circular polarization.  While the PLANCK experiment seems difficult 
to answer whether the GWB is polarized or not, 
post PLANCK experiments dominated by the cosmic-variance may give 
a meaningful constraint on the parity violation in the early universe. 
This result is interesting in the sense that the next-generation 
laser interferometers will also be sensitive to the 
circular polarization mode of primordial gravitational waves  
\cite{Seto:2006hf,Seto:2006dz,Seto:2006ip}. Although, in practice, 
a large value of $\varepsilon$ is required to falsify the possibility 
of an un-polarized GWB, combined results of the two different measurements 
lead to a valuable implication of the physics beyond standard 
inflationary predictions.

\par

\ack
We would like to thank Eiichiro Komatsu for many helpful comments and 
discussions. We also thank Yasushi Suto, Shinji Mukohyama, Jun'ichi Yokoyama, 
Kazuhiro Yahata, Shun'ichiro Kinoshita, Takahiro  Nishimichi, Yudai Suwa, 
and Erik Reese for useful discussions. K. I acknowledges the support from the Japan 
Society for Promotion of Science (JSPS) research fellows. 
A.T is supported by a Grant-in-Aid for
Scientific Research from the JSPS (No.18740132).

\appendix
\setcounter{section}{0}

\section{CMB power spectra from tensor perturbation} 
\label{sec:power_spectra}

 In this appendix, we summarize the explicit form of the CMB power spectra 
for tensor modes.  First write down the CMB power spectra as \cite{Zaldarriaga:1996xe}: 
\ba
&& C^{{\rm XY}(t)}_{\ell}=(4\pi)^{2}\int k^{2}dk P^{t}(k)
\Delta^{t}_{{\rm X}\ell}(k)\Delta^{t}_{{\rm Y}\ell}(k)\ ;
\quad \mbox{(X,Y =T,E,B)},
\label{eq:A1}
\ea
where the photon transfer functions $\Delta^t_{{\rm X}\ell}(k)$ are the 
multipole moment of the function $\widetilde{\Delta}^{t}_{X}(k,\mu)$ (see below) 
and their explicit expressions are given by the integral form:  
\ba
&&\quad\quad\quad \Delta^{t}_{{\rm T}\ell}(k)= \sqrt{\frac{(\ell+2)!}{(\ell-2)!}}
\int^{\eta_{0}}_{0}d\eta (-h'e^{-\tau}+g\Psi)P_{{\rm T}\ell}(x)\ ,
\label{eq:A2}
\\
&&\quad\quad\quad \Delta^{t}_{{\rm E}\ell}(k)= 
\int^{\eta_{0}}_{0}d\eta (-g\Psi) P_{{\rm E}\ell}(x)\ ,
\label{eq:A3}
\\
&&\quad\quad\quad \Delta^{t}_{{\rm B}\ell}(k)= 
\int^{\eta_{0}}_{0}d\eta (-g\Psi) P_{{\rm B}\ell}(x)\ ,
\label{eq:A4}
\ea
with the quantity $g$ being the visibility function defined by 
\ba
g(\eta)=\tau' \,e^{-\tau}. 
\label{eq:visibility}
\ea
Here, $\tau$ is the optical depth for Thomson scattering between a given conformal 
time $\eta$  and the present time $\eta_0$,  
the quantity $h$ is the amplitude of the gravitational waves and 
$x\equiv k(\eta_{0}-\eta)$. The prime denotes the derivative with respect to 
the conformal time $\eta$, and 
the subscript $t$ indicates the contribution from tensor modes.  
In the above expressions, 
the functions $\Psi$ is the source function for radiative transfer of photon 
and $P_{{\rm E,B}\ell}$ are the projection factors for each polarization mode of photon, 
given by 
\ba
 \Psi & \equiv & \frac{1}{10}\widetilde{\Delta}^{t}_{{\rm T}0}+\frac{1}{7}
\widetilde{\Delta}^{t}_{{\rm T}2}+\frac{3}{70}\widetilde{\Delta}^{t}_{{\rm T}4}
-\frac{3}{5}\widetilde{\Delta}^{t}_{{\rm P}0}+\frac{6}{7}\widetilde{\Delta}^{t}_{{\rm P}2}
-\frac{3}{70}\widetilde{\Delta}^{t}_{{\rm P}4}\ ,
\label{eq:A5}
\\
 P_{{\rm T}\ell}(x)&\equiv& \frac{j_{\ell}(x)}{x^{2}}\ ,
\label{eq:A6a}
\\
 P_{{\rm E}\ell}(x)&\equiv& -j_{\ell}(x)+\partial^{2}_{x}j_{\ell}(x)
 +\frac{2j_{\ell}(x)}{x^2}+
\frac{4\partial_{x}j_{\ell}(x)}{x}\ ,
\label{eq:A6}
\\
 P_{{\rm B}\ell}(x)&\equiv& 2\partial_{x}j_{\ell}(x)+\frac{4j_{\ell}(x)}{x}\ ,
\label{eq:A7}
\ea
 where $j_{\ell}(x)$ is the $\ell$-th Bessel function.
The expressions (\ref{eq:A1})--(\ref{eq:A7}) are basically derived from the 
Boltzmann equations for photon's radiative transfer. To derive equations, first note
that the quantities $\widetilde{\Delta}^{t}_{X}(k,\mu)$ are the solutions 
of the Boltzmann equation, which are formally written as the line-of-sight 
integral form:   
\ba
 \widetilde{\Delta}^{t}_{\rm T}(k,\mu)  
=\int_0^{\eta_0} d\eta\,e^{-i\,x\,\mu}(-h'\,e^{-\tau}+g\Psi),
\label{eq:A8}
\\
 \widetilde{\Delta}^{t}_{\rm E}(k,\mu)  =
\{-12+x^{2}(1-\partial_{x}^{2})-8x\partial_{x}\}
\widetilde{\Delta}^{t}_{\rm P}(k,\mu),
\label{eq:A9}
\\
 \tilde{\Delta}^{t}_{\rm B}(k,\mu) = \{8x+2x^{2}\partial^{2}_{x}\}
\tilde{\Delta}^{t}_{\rm P}(k,\mu)
\label{eq:A10}
\ea
with the quantity $\widetilde{\Delta}^{t}_{\rm P}(k,\mu)$ being 
\ba
\widetilde{\Delta}^{t}_{\rm P}(k,\mu)= \int^{\eta_{0}}_{0}d\eta \,
e^{-i\,x\,\mu}(-g\Psi).
\label{eq:A11}
\ea
The quantities $\widetilde{\Delta}^{t}_{X}(k,\mu)$ are related to 
the direct observables of the temperature and the polarization maps, $X^{t}(\hat{n})$.  
Writing the projected maps as $X^{t}(\hat{n})=\int d^{3}k\Delta^{t}_{X}(k,\hat{n})$, 
the relation between $\Delta^{t}_{X}(k,\hat{n})$ and $\widetilde{\Delta}^{t}_{X}(k,\mu)$ 
are given by 
\ba
 \Delta^{t}_{\rm T}({\bf k},\hat{n})  = 
[(1-\mu^{2})e^{2i\phi}\xi^{\rm R}({\bf k})+
(1-\mu^{2})e^{-2i\phi}\xi^{\rm L}({\bf k})]\widetilde{\Delta}^{t}_{\rm T}(k,\mu)\ ,
\label{eq:A12}
\\
 \Delta^{t}_{\rm E}({\bf k},\hat{n})  =  
[(1-\mu^{2})e^{2i\phi}\xi^{\rm R}({\bf k})+
(1-\mu^{2})e^{-2i\phi}\xi^{\rm L}({\bf k})]\widetilde{\Delta}^{t}_{\rm E}(k,\mu)\ ,
\label{eq:A13}
\\
 \Delta^{t}_{\rm B}({\bf k},\hat{n})  =  
[-(1-\mu^{2})e^{2i\phi}\xi^{\rm R}({\bf k})+(1-\mu^{2})
e^{-2i\phi}\xi^{\rm L}({\bf k})]\widetilde{\Delta}^{t}_{\rm B}(k,\mu)\ . 
\label{eq:A14}
\ea
Here, the variables, $\xi^{\rm L,R}({\bf k})$, are the independent random variables 
characterizing the statistical properties of the GWB. In this paper, we assume that
\ba
 \langle \xi^{{\rm L}*}({\bf k})\xi^{\rm L}({\bf k}') \rangle = 
\delta({\bf k}-{\bf k}')\,P^{t{\rm L}}(k)\ ,
\label{eq:A15}
\\
 \langle \xi^{{\rm R}*}({\bf k})\xi^{\rm R}({\bf k}') \rangle = 
\delta({\bf k}-{\bf k}')\,P^{t{\rm R}}(k)\ ,
\label{eq:A16}
\\
 \langle \xi^{{\rm L}*}({\bf k})\xi^{\rm R}({\bf k}') \rangle = 0\ .
\label{eq:A17}
\ea

Starting from the line-of-sight integral solutions of the Boltzmann equation 
(\ref{eq:A8})--(\ref{eq:A10}) and using the relations (\ref{eq:A12})--(\ref{eq:A17}),    
one can derive the expressions for the CMB power spectra (\ref{eq:A1})--(\ref{eq:A2})    
with help of the definition (\ref{eq:definition of CMB power spectra}) 
and the multipole expansion of the anisotropies on a projected sky: 
\be
  a^{\rm X}_{\ell m} = \int d\Omega \,\,Y^{*}_{\ell m}(\hat{n})
\int d^{3}{\bf k}\,\, \Delta^{t}_{\rm X}(\eta_{0},{\bf k},\hat{n})\ .
\label{eq:multipole_alm}
\ee
For details of the derivation, the readers may refer to 
Refs.\cite{Zaldarriaga:1996xe,Lin:2004xy}.

 \section{Linear polarization of the GWB and CMB power spectra} 
 \label{sec:linear-polarized}

In this paper, we have mainly focused on the detectability of circularly 
polarized GWB. Here, we briefly discuss the measurability of a linearly 
polarized GWB.

Let us recall that the circularly polarized states of gravitational 
waves are related to the linearly polarized states as 
$ h^{L}=(h^{+}+ih^{\times})/\sqrt{2}$ and $ h^{R}=(h^{+}-ih^{\times})/\sqrt{2}$. 
Using this relationship, the power spectra of linearly polarized GWB can be 
rewritten with 
\ba
 P^{t+}(k)& = & \langle \xi^{+*}\xi^{+} \rangle 
\nonumber\\
 &=& \Bigl\langle(\xi^{{\rm L}*}+\xi^{{\rm R}*})(\xi^{\rm L}+\xi^{\rm R}) \Bigr\rangle /2 
\nonumber\\
 &=& P^{t{\rm C}}(k)+\{P^{t{\rm R}}(k)+P^{t{\rm L}}(k)\}/2 
\nonumber\\
 &=& P^{t{\rm C}}(k)+P^{t}(k)/2\ ,
\\
 P^{t\times}(k)& = &\Bigl\langle \xi^{\times *}\xi^{\times} \Bigr\rangle 
\nonumber\\
 &=& \Bigl\langle (\xi^{{\rm L}*}-\xi^{{\rm R}*})(\xi^{\rm L}-\xi^{\rm R}) \rangle /2 
\nonumber\\
 &=& -P^{t{\rm C}}(k)+\{P^{t{\rm R}}(k)+P^{t{\rm L}}(k)\}/2 
\nonumber\\
 &=& -P^{t{\rm C}}(k)+P^{t}(k)/2\ ,
\ea
where we have defined the cross power spectrum between 
left- and right-handed polarized states by 
$P^{t{\rm C}}(k)\equiv\langle \xi^{{\rm L}*}\xi^{\rm R}+\xi^{{\rm R}*}
\xi^{\rm L} \rangle/2$. The above expressions readily imply that the linearly 
polarized GWB comes from the non-vanishing contribution of cross power spectrum 
$P^{t{\rm C}}(k)$. Thus, the crucial question is whether the cross power spectrum 
$P^{t{\rm C}}(k)$ is measurable or not.

To clarify this, consider the TT-mode power spectra. Following the definition 
(\ref{eq:definition of CMB power spectra}), we have 
\ba
 C^{\mathrm{TT}(t)}_{\ell}&=&\frac{1}{2\ell+1}\sum_{m}
\langle a^{{\rm T}*}_{\ell m}a^{\rm T}_{\ell m}\rangle 
\nonumber\\
 &=& \frac{1}{2\ell+1}\sum_{m} \int d\Omega'\int d\Omega \int d^{3}{\bf k}' 
\int d^{3}{\bf k} \,\,
Y_{\ell m}(\hat{n}')Y^{*}_{\ell m}(\hat{n})\,\,\nonumber\\
 && \; \; \times \Bigl\langle\,\Delta^{t*}_{{\rm T}}(\eta_{0},{\bf k}',\hat{n}')
\Delta^{t}_{{\rm T}}(\eta_{0},{\bf k},\hat{n})\,\Bigr\rangle. 
\ea
from equation (\ref{eq:multipole_alm}). 
The substitution of equation (\ref{eq:A12}) into the above expression leads to 
\ba
 C^{\mathrm{TT}(t)}_{\ell}
 &=& \frac{1}{2\ell+1}\sum_{m}\, \int d\Omega'\int d\Omega 
\int d^{3}{\bf k}' \int d^{3}{\bf k}\,\, Y_{\ell m}(\hat{n}')Y^{*}_{\ell m}(\hat{n}) 
\nonumber
\\
 && \; \; \times \Bigl\langle
(1-{\mu'}^{2})\left\{e^{-2i\phi'}\xi^{{\rm R}*}({\bf k}')+
e^{2i\phi}\xi^{{\rm L}*}({\bf k}')\right\}\widetilde{\Delta}^{t*}_{{\rm T}}(\eta_{0},k')
\nonumber\\
 && \; \; \; \; \; \; \times (1-{\mu}^{2})\left\{e^{2i\phi}\xi^{\rm R}({\bf k})+
e^{-2i\phi}\xi^{\rm L}({\bf k})\right\}\widetilde{\Delta}^{t}_{{\rm T}}(\eta_{0},k)
\Bigr\rangle 
\nonumber\\
 &=& \frac{1}{2\ell+1}\sum_{m}\, \int d\Omega'\int d\Omega \int d^{3}{\bf k}' 
\int d^{3}{\bf k}\,\, Y_{\ell m}(\hat{n}')Y^{*}_{\ell m}(\hat{n}) 
\nonumber\\
 && \; \; \times (1-{\mu'}^{2})(1-{\mu}^{2})\widetilde{\Delta}^{t*}_{{\rm T}}(\eta_{0},k')\widetilde{\Delta}^{t}_{{\rm T}}(\eta_{0},k)
\nonumber\\
 && \; \; \times \left[(e^{2i\phi}e^{-2i\phi'}+e^{-2i\phi}e^{-2i\phi'})
P^{t}(k)\delta({\bf k}-{\bf k}')/2\right.
\nonumber\\
 && \; \; \; + \left. \Bigl\langle e^{-2i\phi'}e^{-2i\phi}\xi^{R*}({\bf k}')
\xi^{L}({\bf k})+e^{2i\phi'}e^{2i\phi}\xi^{L*}({\bf k}')\xi^{R}({\bf k})\Bigr\rangle 
\right]\ .
\label{eq:A1_1}
\ea
In the last line of equation (\ref{eq:A1_1}), there appears the 
cross-correlation of the ensemble between $\xi^{L}$ and $\xi^{R}$, which represents 
the contribution from the linearly polarized GWB. However, further proceeding 
to the integral over the azimuthal angle $\phi$, it turns out that this term becomes 
vanishing. Because of $Y_{\ell m}\propto e^{im\phi}$, the relevant part of 
the integral over $\phi$ can be written as 
\be
 \int^{2\pi}_{0} d\phi e^{\pm 2i\phi}e^{-im\phi}=2\pi\delta_{m\pm 2}\ , 
\ee
which thus leads to 
\be
 \left(\int^{2\pi}_{0} d\phi' e^{2i\phi'}e^{im\phi'}\right)\left(\int^{2\pi}_{0} 
d\phi e^{2i\phi}e^{-im\phi}\right)=(2\pi)^{2}\delta_{m2}\delta_{m-2}=0\ .
\ee
Hence, linearly polarized GWB is shown to be null effect 
on the TT-mode power spectrum.  
Note that similar argument does hold for the other power spectra, since all of the 
photon transfer functions $\Delta_{\rm X}^t$ are written by a 
linear combination of $e^{2i\,\phi}\xi^{\rm R}$ and $e^{-2i\,\phi}\xi^{\rm L}$ 
(see Eqs.(\ref{eq:A12}), (\ref{eq:A13}) and (\ref{eq:A14})). Thus, the 
cross correlation term always has the factor $e^{-2i\phi'}e^{-2i\phi}$ or 
$e^{2i\phi'}e^{2i\phi}$, which finally vanishes after the integration over 
the azimuthal angle.

 \section{Characteristic features of TB- and EB-mode power spectra}
 \label{sec:TBEB}
 In this appendix, we discuss the details in characteristic features 
of TB- and EB-mode power spectra originating from circularly polarized GWB.\par
 \S \ref{sec:general_feature} reveals that while the TB-mode power spectrum has 
one zero-crossing point around $\ell\sim 70$, the EB-mode power spectrum has many 
zero-crossing points with tiny amplitudes. These features are mainly attributed to 
the projection factors in photon's transfer function, 
$\Delta^{t}_{\rm{T,E,B}\ell}$ (see (\ref{eq:A6a})-(\ref{eq:A7})).\par
\ba
P_{{\rm T}\ell}(x)&\equiv& \frac{j_{\ell}(x)}{x^{2}}\ ,\nonumber
\\
 P_{{\rm E}\ell}(x)&\equiv& -j_{\ell}(x)+\partial^{2}_{x}j_{\ell}(x)
 +\frac{2j_{\ell}(x)}{x^2}+
\frac{4\partial_{x}j_{\ell}(x)}{x}\ ,\nonumber
\\
 P_{{\rm B}\ell}(x)&\equiv& 2\partial_{x}j_{\ell}(x)+\frac{4j_{\ell}(x)}{x}\ ,
\nonumber
\ea
 
\begin{figure}
\begin{center}
\includegraphics[width=0.5\textwidth]{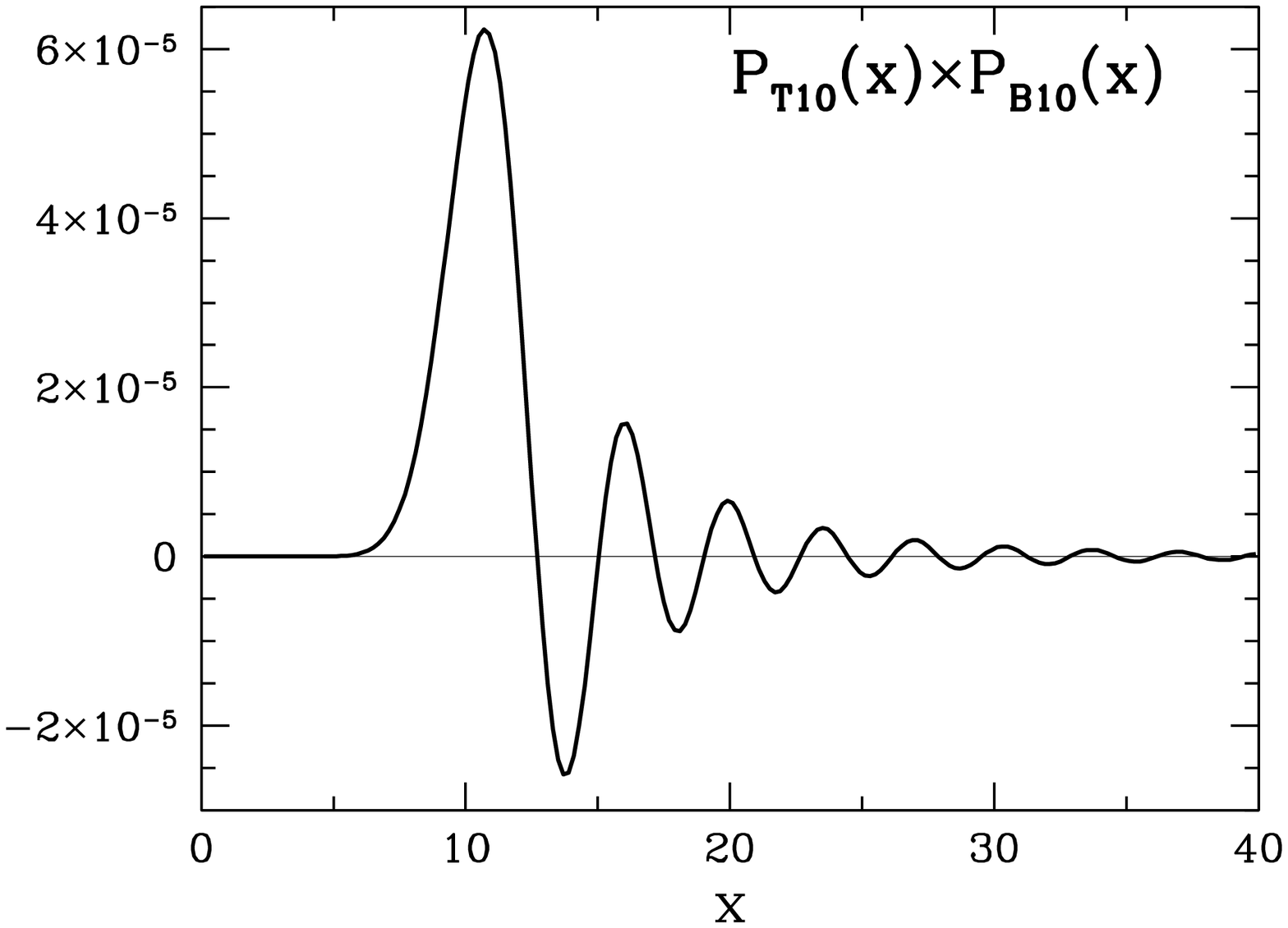}
\includegraphics[width=0.5\textwidth]{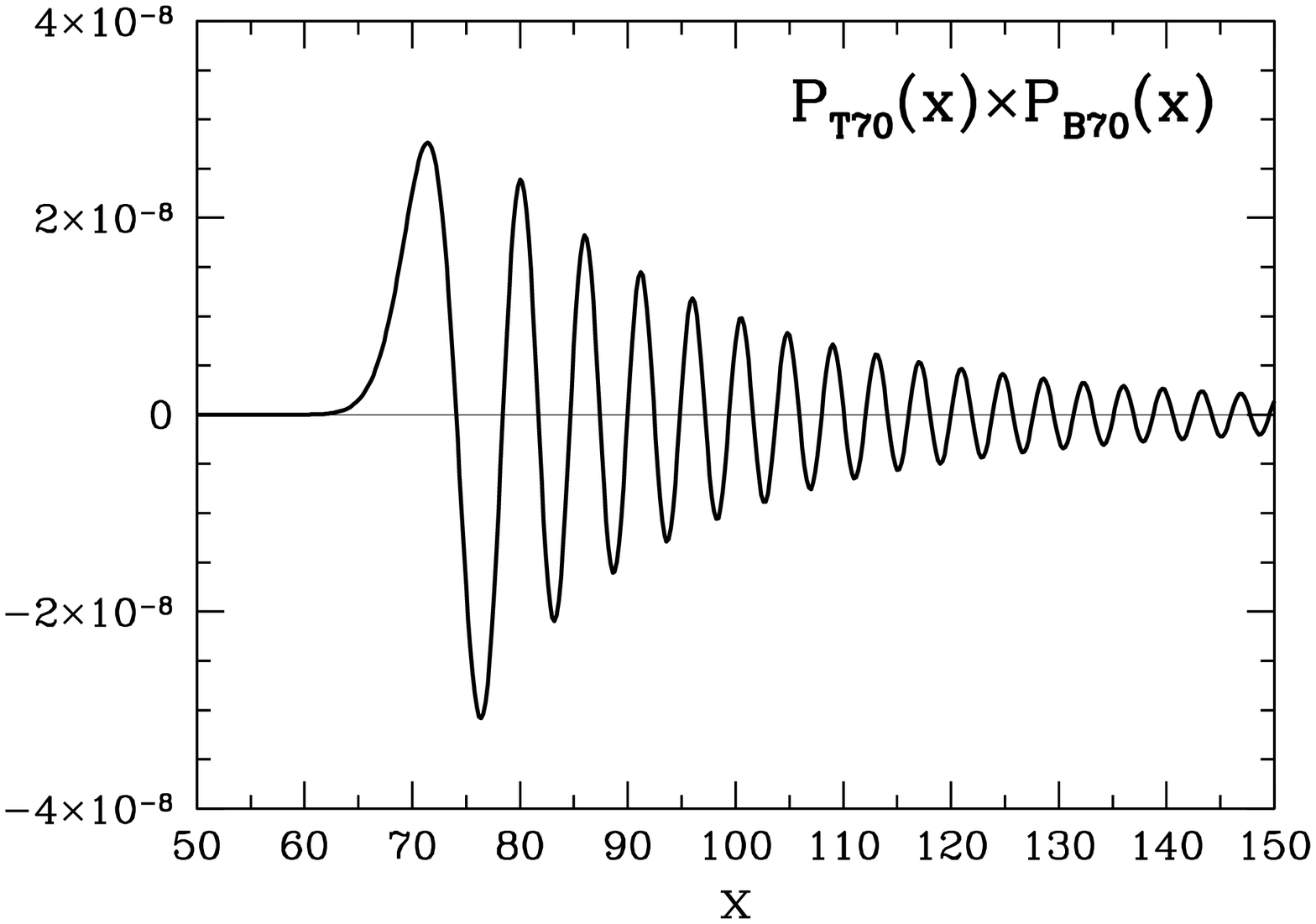}
\includegraphics[width=0.5\textwidth]{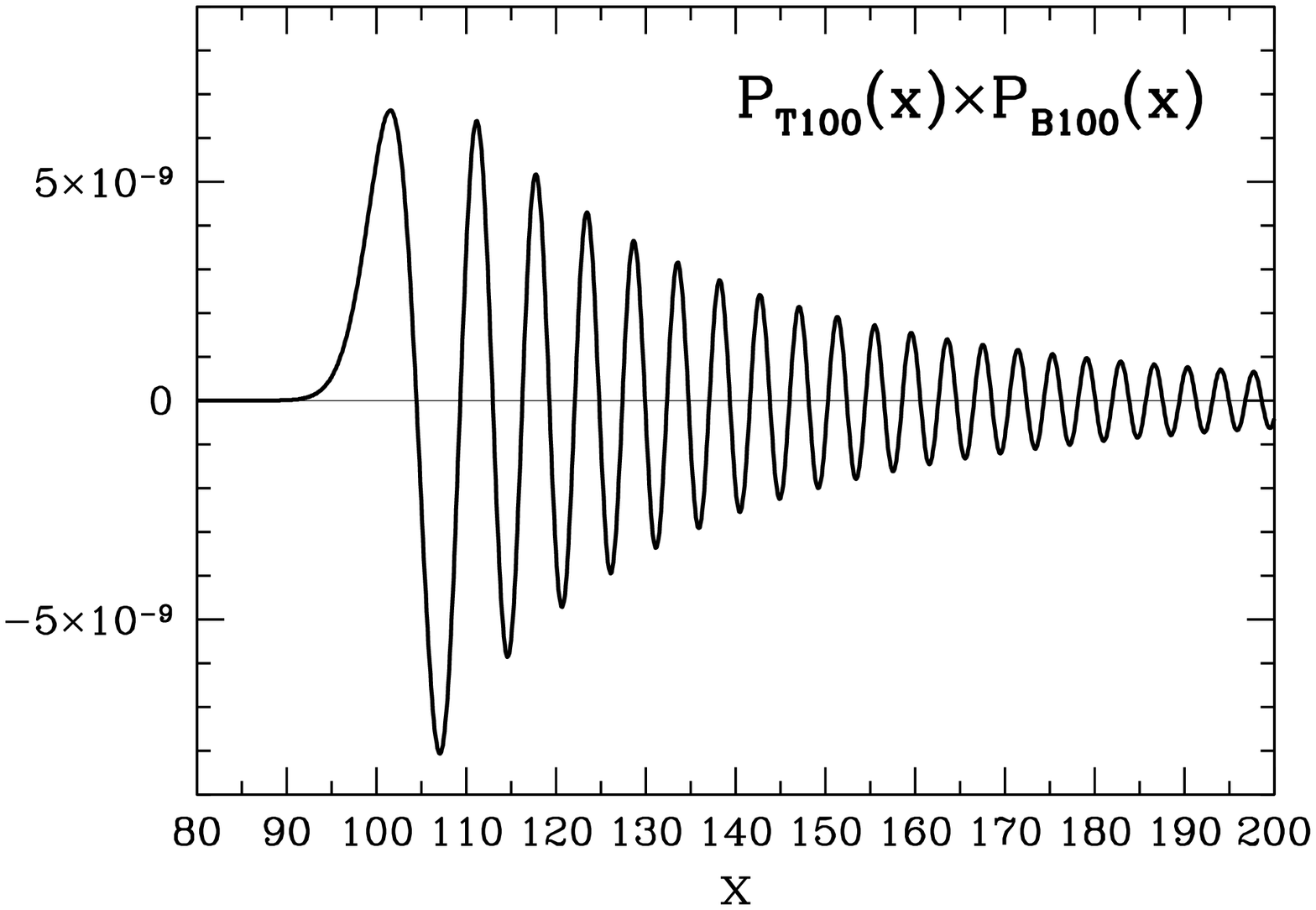}
\end{center}
\caption{The projection factor of TB-mode spectrum, 
$P_{{\rm T}\ell}(x) \times P_{{\rm B}\ell}(x)$, as function of 
$x=k(\eta_0-\eta)$. The {\it top}, {\it middle}, and {\it bottom} 
panels represent the results with multipoles $\ell=10$, $\ell=70$, and $\ell=100$, 
respectively.}
\label{fig.A1}
\end{figure}

Let us first consider the TB-mode spectrum, in which there appears the 
projection factors, $P_{{\rm T}\ell}(x)\times P_{{\rm B}\ell}(x)$, 
in the kernel of the integral (\ref{eq:A1}). Figure \ref{fig.A1} shows the 
function, $P_{{\rm T}\ell}(x)\times P_{{\rm B}\ell}(x)$, as function of 
$x=k(\eta_0-\eta)$ for various multipoles with  
$\ell=10$ ({\it top}), $70$ ({\it middle}) and $100$ ({\it bottom}). 
The function starts to oscillate around $x\sim\ell$ and it asymptotically decays as 
$x^{-4}$. Thus, the main contribution to the integral in equation (\ref{eq:A1}) 
comes from the first several peaks in the oscillations and many oscillations at large 
$x$ are almost canceled out. Just focusing on the first part around $x\sim\ell$, we 
find that the positive part of the oscillating amplitudes has relatively larger values 
for lower-multipoles, while the amplitude at higher-multipoles has slightly large 
negative amplitudes. Eventually, the values of the positive and negative 
amplitudes become comparable at the multipole $\ell\sim70$. These behaviors 
quantitatively explain the shape of the TB-mode power spectrum. 
\par
\begin{figure}
\begin{center}
\includegraphics[width=0.6\textwidth]{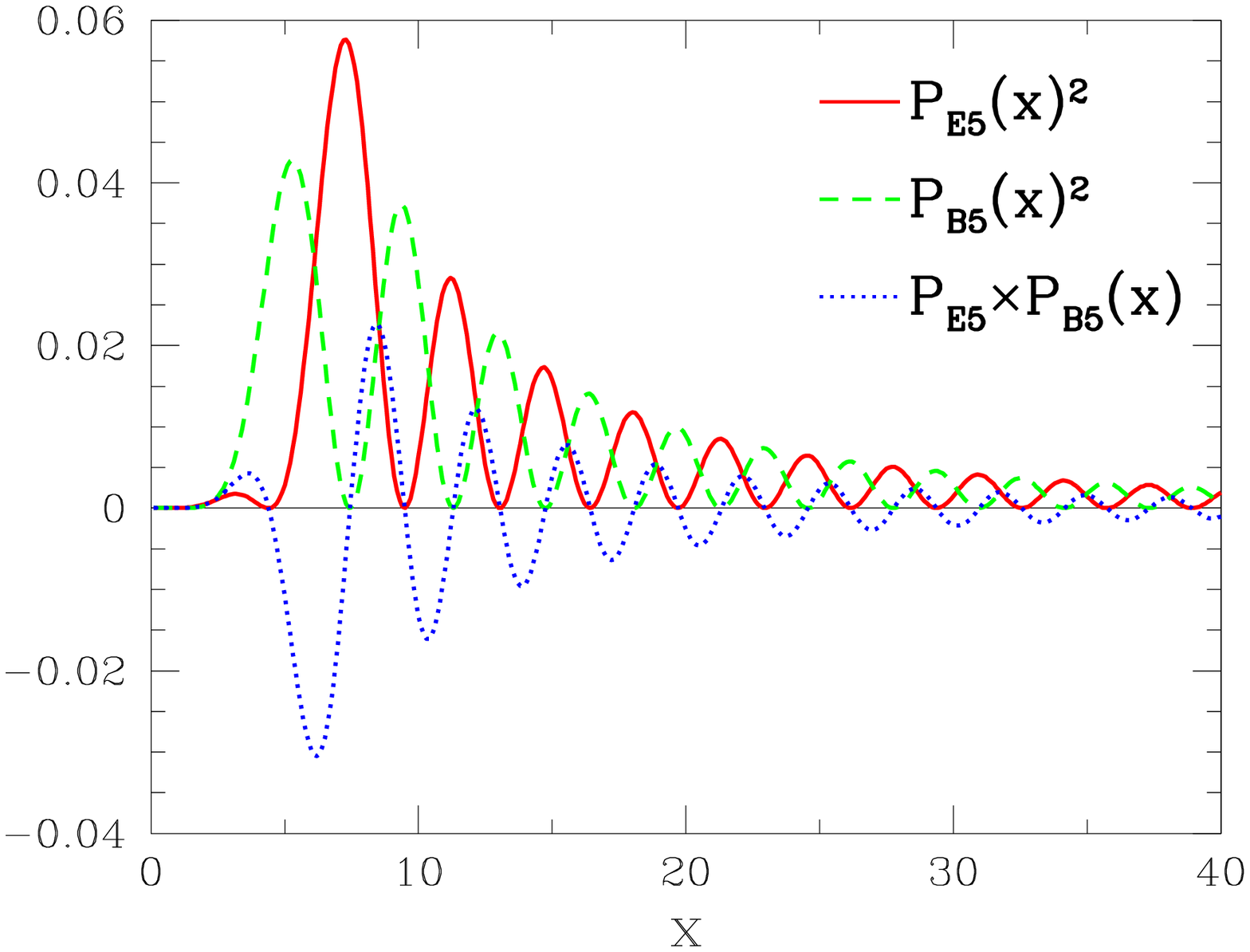}
\end{center}
\caption{The projection factors of EE-, BB- and EB-mode power spectra at the 
multipole $\ell=5$: 
$P_{{\rm E}\ell}(x)^{2}$({\it red, solid}), 
$P_{{\rm B}\ell}(x)^{2}$({\it green, short-dash}), and 
$P_{{\rm E}\ell}(x)\times P_{B\ell}(x)$({\it blue, dot}).}
\label{fig.A2}
\end{figure}

Similarly, tiny amplitude of the EB-mode spectrum is also explained by the projection 
factor, $P_{{\rm E}\ell}(x)$ and $P_{{\rm B}\ell}(x)$. In Figure \ref{fig.A2}, we 
plot the projection factor of EB-mode power spectrum, 
$P_{{\rm E}\ell}(x)\times P_{{\rm B}\ell}(x)$ ({\it blue dot}), 
together with those of the BB- and EE-mode spectra, 
$P_{{\rm B}\ell}(x)^2$ ({\it green, short-dashed}) and 
$P_{{\rm E}\ell}(x)^2$ ({\it red, solid}). 
The oscillation of the projection factor $P_{{\rm B}\ell}(x)^2$ is $\pi/2$ out of phase 
with corresponding one of EE-mode power spectrum. Thus, 
the amplitude of the product, 
$P_{{\rm E}\ell}(x)\times P_{{\rm B}\ell}(x)$, is degraded as a result of phase 
cancellation. Hence, the amplitude of EB-mode power spectrum becomes much smaller than 
those of the EE- and BB-mode spectra.


 \section{Weak lensing effect on CMB power spectra}
\label{sec:weak_lensing}

 Here, we derive the expression for the lensed TB-mode power spectrum following 
the discussion in Ref.\cite{Zaldarriaga:1998ar}. The gravitational lensing effect 
appears as the angular excursion of the photon path. Since the lensing effect 
is only relevant at the small angular scales in the CMB, we consider the small 
scale limit. In terms of Fourier components we have following the 
expressions for the Stokes parameters:
\ba
 T(\vec{\theta})&=&\widetilde{T}(\vec{\theta}+\delta\vec{\theta})=
\int\frac{d^{2}\vec{\ell}}{(2\pi)^{2}}e^{i\vec{\ell}
\cdot(\vec{\theta}+\delta\vec{\theta})}T(\vec{\ell})\ ,
\nonumber\\
 Q(\vec{\theta})&=&\widetilde{Q}(\vec{\theta}+\delta\vec{\theta})=
\int\frac{d^{2}\vec{\ell}}{(2\pi)^{2}}e^{i\vec{\ell}
\cdot(\vec{\theta}+\delta\vec{\theta})}Q(\vec{\ell})\ ,
\nonumber\\
 U(\vec{\theta})&=&\widetilde{U}(\vec{\theta}+\delta\vec{\theta})=
\int\frac{d^{2}\vec{\ell}}{(2\pi)^{2}}e^{i\vec{\ell}
\cdot(\vec{\theta}+\delta\vec{\theta})}U(\vec{\ell})\ ,
\label{eq:weak_lensing_Stokes}
\ea
 where $\widetilde{X}$ describes the unlensed X.

 The polarization parameter $Q$ and $U$ can be expressed with $E$ and $B$ as:
\ba
 Q(\vec{\ell})&=& E(\vec{\ell})\cos2\phi_{\ell}-B(\vec{\ell})\sin2\phi_{\ell}\ ,
\nonumber\\
 U(\vec{\ell})&=& E(\vec{\ell})\sin2\phi_{\ell}+B(\vec{\ell})\cos2\phi_{\ell}\ ,
\label{eq:QEvsEB}
\ea
where $\phi_{\ell}$ is the azimuthal angle. The ensemble average of each 
Fourier components becomes 
\be
 \langle \widetilde{X}(\vec{\ell})\widetilde{Y}(\vec{\ell}')\rangle=(2\pi)^{2}
C^{\widetilde{{\rm X}}\widetilde{{\rm Y}}}_{\ell}\delta(\vec{\ell}-\vec{\ell}') 
\label{eq:ansemble}
\ee
with $\widetilde{X},\widetilde{Y}=\widetilde{T},\widetilde{E}$ and $\widetilde{B}$. 
Using equations (\ref{eq:weak_lensing_Stokes}), (\ref{eq:QEvsEB}) and (\ref{eq:ansemble}), 
cross correlation functions $ C_{{\rm TQ}}$ and $ C_{{\rm TU}}$ are expressed as 
\ba
 C_{{\rm TQ}}(\theta)&=& \int\frac{d^{2}\vec{\ell}}{(2\pi)^{2}}
e^{i\ell\theta\cos\phi_{\ell}}\langle e^{i\vec{\ell}
\cdot(\delta\vec{\theta}_{A}-\delta\vec{\theta}_{B})}\rangle 
[C^{\widetilde{{\rm T}}\widetilde{{\rm E}}}_{\ell}
\cos2\phi_{\ell}-C^{\widetilde{{\rm T}}\widetilde{{\rm B}}}_{\ell}\sin2\phi_{\ell}] ,
\nonumber\\
 C_{{\rm TU}}(\theta)&=& \int\frac{d^{2}\vec{\ell}}{(2\pi)^{2}}
e^{i\ell\theta\cos\phi_{\ell}}\langle e^{i\vec{\ell}
\cdot(\delta\vec{\theta}_{A}-\delta\vec{\theta}_{B})}\rangle 
[C^{\widetilde{{\rm T}}\widetilde{{\rm E}}}_{\ell}
\sin2\phi_{\ell}-C^{\widetilde{{\rm T}}\widetilde{{\rm B}}}_{\ell}\cos2\phi_{\ell}] ,
\label{eq:TQTUvsTETB}
\ea
where we set $\theta$ as $\theta\equiv \theta_{A}-\theta_{B}$. 
The above expressions still possess the ensemble average, which represents the 
statistical average over the photon excursions caused by the lensing effect. 
In the weak lensing limit, the angular excursion is approximately 
described by the random Gaussian distribution with small dispersion. We have 
(e.g., Ref.\cite{Seljak:1995ve}):
\ba
 \langle e^{i\vec{\ell}\cdot(\delta\vec{\theta}_{A}-\delta\vec{\theta}_{B})}
\rangle&=& \exp\left\{-\frac{\ell^{2}}{2}[\sigma_{0}^{2}(\theta)+
\cos(2\phi_{\ell})\sigma_{2}^{2}(\theta)]\right\}
\nonumber\\
 &\simeq& 1-\frac{\ell^{2}}{2}[\sigma_{0}^{2}(\theta)+\cos(2\phi_{\ell})
\sigma_{2}^{2}(\theta)]\ .
\label{eq:correlation of excursion}
\ea
The functions, $\sigma^{2}_{0}$ and $\sigma^{2}_{2}$, characterize the rms 
fluctuations of the photon path (see Ref.\cite{Zaldarriaga:1998ar} for 
their explicit expressions). 
Substituting the relation (\ref{eq:correlation of excursion}) into 
(\ref{eq:TQTUvsTETB}) and integrating over the azimuthal angle $\phi_{\ell}$, 
we obtain
\ba
 &&C_{{\rm TQ}}(\theta)=
-\int\frac{\ell d\ell}{2\pi}C^{\widetilde{{\rm T}}\widetilde{{\rm E}}}_{\ell}
\left[J_{2}(\ell\theta)\left\{1-\frac{\ell^{2}\sigma_{0}^{2}(\theta)}{2}\right\}
+\frac{\ell^{2}\sigma_{2}^{2}(\theta)}{4}\left\{J_{0}(\ell\theta)+
J_{4}(\ell\theta)\right\}\right]\ ,
\nonumber\\
 &&C_{{\rm TU}}(\theta)= 
-\int\frac{\ell d\ell}{2\pi}C^{\widetilde{{\rm T}}\widetilde{{\rm B}}}_{\ell}
\left[J_{2}(\ell\theta)\left\{1-\frac{\ell^{2}\sigma_{0}^{2}(\theta)}{2}\right\}
+\frac{\ell^{2}\sigma_{2}^{2}(\theta)}{4}\left\{J_{0}(\ell\theta)
+J_{4}(\ell\theta)\right\}\right]\ .\nonumber\\
 && 
\ea
The above expressions finally lead to the angular power spectrum of TB mode 
with a help of the relation: 
\be
 C^{{\rm TB}}_{\ell}=-2\pi \int^{\pi}_{0}\theta d\theta C_{\rm {TU}}(\theta)J_{2}(\ell\theta)
\ ,
\ee
The resultant expression becomes
\ba
 C^{{\rm TB}}_{\ell}&=&C^{\widetilde{{\rm T}}\widetilde{{\rm B}}}_{\ell}+\sum_{\ell'}
\mathcal{W}^{\ell'}_{\ell}C^{\widetilde{{\rm T}}\widetilde{{\rm B}}}_{\ell'}\ ,
\\
 \mathcal{W}^{\ell'}_{\ell}&=&\int^{\pi}_{0}\theta d\theta J_{2}(\ell\theta)
\left[-\frac{{\ell'}^{3}}{2}\sigma_{0}^{2}(\theta)J_{2}(\ell'\theta)+
\frac{{\ell'}^{3}}{4}\sigma_{2}^{2}(\theta)\{J_{0}(\ell'\theta)+
J_{4}(\ell'\theta)\}\right]\ .\nonumber\\
&&
\ea
That is, the lensed power spectrum of the TB-mode is generated 
if and only if 
the primary TB-mode exists. No other cross spectra can create 
the lensed TB-mode. 
This may be explained intuitively by a simple symmetry reason; 
the change of 
TE-mode into TB-mode breaks parity which we do not expect from 
weak lensing effect.
Since the transformation matrix $ \mathcal{W}^{\ell'}_{\ell}$ is the oscillating function whose amplitude is basically much less than 
unity \cite{Zaldarriaga:1998ar}, 
the lensing effect on the TB-mode spectrum can be safely neglected 
as long as the primary TB-mode spectrum is generated from the 
tensor-type fluctuations.

\section{Exact form of likelihood function} 
\label{sec:likelihood}

 In this Appendix, we briefly sketch the derivation of the exact likelihood 
function used in \S \ref{sec:Future prospects}. To do this, we first follow 
the simplest case of the likelihood function with temperature anisotropy 
data alone.
The likelihood function for the temperature anisotropies observed 
by a perfect experiment (i.e., noiseless and full-sky observation) 
has the following form:
\be
 \mathcal{L}(\vec{T}|\bar{C}^{\rm TT}_{\ell})\propto 
\frac{1}{\sqrt{|\mathbf{S}|}}
\exp\left[-\frac{\vec{T}^{\rm T}\mathbf{S}^{-1}\vec{T}}{2}\right]\ ,
\ee
 where $\vec{T}$ denotes our temperature map, $\mathbf{S}$ is correlation matrix 
given by $S_{ij}=\sum_{\ell}(2\ell+1)\overline{C}^{\rm TT}_{\ell}\,P_{\ell}
(\hat{n}_{i}\cdot\hat{n}_{j})/(4\pi)$,  where the $P_{\ell}$ are the 
Legendre polynomials 
and $\hat{n}_{i}$ is the pixel position on the map, and $|\mathbf{S}|$ denotes 
determinant of correlation matrix. Expanding the temperature map in spherical 
harmonics: $T(\hat{n})=\sum_{\ell m}a^{\rm T}_{\ell m}Y_{\ell m}(\hat{n})$, 
the likelihood function for each $a^{\rm T}_{\ell m}$ becomes 
\be
 \mathcal{L}(\vec{T}|\overline{C}^{\rm TT}_{\ell})\propto 
\prod_{\ell m}\frac{1}{\sqrt{\bar{C}^{\rm TT}_{\ell}}}
\exp\left(-\frac{|a^{\rm T}_{\ell m}|^{2}}{2\overline{C}^{\rm TT}_{\ell}}\right).
\ee
If we assume that each multipole moment $a^{\rm T}_{\ell m}$ just follows 
the Gaussian statistics with 
variance of $\bar{C}^{\rm TT}_{\ell}$, the above expression can be 
reduced to a$\chi^{2}$-distribution with $(2\ell+1)$ degrees 
of freedom:
\be
 -2\ln \mathcal{L} =\sum_{\ell}\left[-(2\ell-1)\ln 
\widehat{C}^{\rm TT}_{\ell}+(2\ell+1)
\left(\ln \overline{C}^{\rm TT}_{\ell}+
\frac{\widehat{C}^{\rm TT}_{\ell}}{\overline{C}^{\rm TT}_{\ell}}-1\right)\right], 
\label{eq:chi^2}
\ee
where $\widehat{C}^{\rm TT}_{\ell}$ denotes the estimator defined by 
$\widehat{C}^{\rm TT}_{\ell}=\sum_{m}|a^{\rm T}_{\ell m}|^{2}/(2\ell+1)$.
\par

Assuming a uniform prior distribution, the posterior distribution function
is proportional to the likelihood function as a result of Bayes' theorem. 
Thus, the likelihood function  (\ref{eq:chi^2}) can be viewed as the 
posterior distribution function, as a function of the theoretical value, 
$\overline{C}^{\rm TT}_{\ell}$. Then, appropriately normalizing the 
posterior distribution, 
the exact expression of the likelihood function for 
temperature anisotropy data is obtained:  
\be
 -2\ln \mathcal{L} (\overline{C}^{\rm TT}_{\ell})=\sum_{\ell}(2\ell+1)
\left[
\ln\left(\frac{\overline{C}^{\rm TT}_{\ell}}{\widehat{C}^{\rm TT}_{\ell}}\right)+
\frac{\widehat{C}^{\rm TT}_{\ell}}{\overline{C}^{\rm TT}_{\ell}}-1\right]. 
\label{eq:TT_likelihood}
\ee
\par

The above result can be extended to the likelihood functions for the general 
case with temperature and polarization anisotropies. 
Restricting the analysis to the case of the 
temperature and B-mode polarization data, 
the likelihood function becomes 
\be
 \mathcal{L} = \prod_{\ell m}\frac{1}{\sqrt{|\mathbf{C}|}}
\exp\left[-\frac{\vec{d}^{T}\mathbf{C}^{-1}\vec{d}}{2}\right],
\ee
where the vector $\vec{d}$ and the matrix $\mathbf{C}$ are respectively given by 
\ba
 \vec{d}^{\rm T} & = & (a^{\rm T}_{\ell m},a^{\rm B}_{\ell m}),\\
 \mathbf{C} & = & \left(
 \begin{array}{cc}
 \overline{C}^{\rm TT}_{\ell} & \overline{C}^{\rm TB}_{\ell}\\
 \overline{C}^{\rm TB}_{\ell} & \overline{C}^{\rm BB}_{\ell}
 \end{array}
 \right).
\ea
Then, just repeating the same procedure as presented above, 
we obtain the likelihood function: 
\ba
 -2\ln \mathcal{L} & = & \sum_{\ell}(2\ell+1)
\left\{\ln\left(\frac{\overline{C}^{\rm TT}_{\ell}\overline{C}^{\rm BB}
_{\ell}-(\overline{C}^{\rm TB}_{\ell})^{2}}
{\widehat{C}^{\rm TT}_{\ell}\widehat{C}^{\rm BB}_{\ell}-
(\widehat{C}^{\rm TB}_{\ell})^{2}}\right)\right.
\nonumber\\
&& \; \; \; 
\left.+\,\,\frac{\widehat{C}^{\rm TT}_{\ell}\overline{C}^{\rm BB}_{\ell}
+\overline{C}^{\rm TT}_{\ell}\widehat{C}^{\rm BB}_{\ell}
-2\overline{C}^{\rm TB}_{\ell}
\widehat{C}^{\rm TB}_{\ell}}{\overline{C}^{\rm TT}_{\ell}
\overline{C}^{\rm BB}_{\ell}-(\overline{C}^{\rm TB}_{\ell})^{2}}-2\right\}. 
\label{eq:exact_likelihood_2}
\ea

\section*{Reference}


\begin{thebibliography}{99}
\bibitem{Bock:2006yf}
  J.~Bock {\it et al.},
  arXiv:astro-ph/0604101.

\bibitem{Verde:2005ff}
  L.~Verde, H.~Peiris and R.~Jimenez,
  JCAP {\bf 0601}, 019 (2006)
  [arXiv:astro-ph/0506036].

\bibitem{Amblard:2006ef}
  A.~Amblard, A.~Cooray and M.~Kaplinghat,
  Phys.\ Rev.\  D {\bf 75}, 083508 (2007)
  [arXiv:astro-ph/0610829].

\bibitem{Cooray:2005xr}
  A.~Cooray,
  Mod.\ Phys.\ Lett. {\bf 20} 2503 (2005)
  [arXiv:astro-ph/0503118].

\bibitem{Smith:2005mm}
  T.~L.~Smith, M.~Kamionkowski and A.~Cooray,
  Phys.\ Rev.\  D {\bf 73}, 023504 (2006)
  [arXiv:astro-ph/0506422].

\bibitem{Smith:2006xf}
  T.~L.~Smith, H.~V.~Peiris and A.~Cooray,
  Phys.\ Rev.\  D {\bf 73}, 123503 (2006)
  [arXiv:astro-ph/0602137].

\bibitem{Kudoh:2005a}
  H.~Kudoh, and A.~Taruya, T.~Hiramatsu and Y.~Himemoto, 
  Phys.\ Rev.\ D {\bf 73}, 064006 (2006)
  [arXiv:gr-qc/0511145].

\bibitem{BBO} E.~Phinney et al, NASA Mission Concept Study (2003).

\bibitem{Seto:2001qf}
  N.~Seto, S.~Kawamura and T.~Nakamura,
  Phys.\ Rev.\ Lett.\  {\bf 87}, 221103 (2001)
  [arXiv:astro-ph/0108011].

\bibitem{2006CQGra..23S.125K}
  S.~Kawamura, et al.,  
  Class.\ Quant.\ Grav.\ {\bf 23}, 125 (2006)

\bibitem{Green84} M.~Green and J.~Shwarz
  Phys.\ Lett.\ B {\bf 149}, 117 (1984) .

\bibitem{Witten84} E.~Witten 
  Phys.\ Lett.\ B {\bf 149}, 351 (1984) .

\bibitem{Lue:1998mq}
  A.~Lue, L.~M.~Wang and M.~Kamionkowski,
  Phys.\ Rev.\ Lett.\  {\bf 83}, 1506 (1999)
  [arXiv:astro-ph/9812088].

\bibitem{Choi:1999zy}
  K.~Choi, J.~c.~Hwang and K.~W.~Hwang,
  Phys.\ Rev.\  D {\bf 61}, 084026 (2000)
  [arXiv:hep-ph/9907244].

\bibitem{Jackiw:2003pm}
  R.~Jackiw and S.~Y.~Pi,
  Phys.\ Rev.\  D {\bf 68}, 104012 (2003)
  [arXiv:gr-qc/0308071].

\bibitem{Lyth:2005jf}
  D.~H.~Lyth, C.~Quimbay and Y.~Rodriguez,
  JHEP {\bf 0503}, 016 (2005)
  [arXiv:hep-th/0501153].

\bibitem{Alexander:2004wk}
  S.~Alexander and J.~Martin,
  Phys.\ Rev.\  D {\bf 71}, 063526 (2005)
  [arXiv:hep-th/0410230].

\bibitem{Alexander06_1}
  S.~Alexander, M.~Peskin and M.~Sheikh-Jabbari
  Phys.\ Rev.\ Lett. {\bf 96}, 081301 (2006) .

\bibitem{Alexander06_2}
  S.~Alexander and S.~Gates Jr. 
  JCAP {\bf 06}, 018 (2006) .

\bibitem{Satoh:2007gn}
  M.~Satoh, S.~Kanno and J.~Soda,
  arXiv:0706.3585 [astro-ph].

\bibitem{Antoniadis:1993jc}
  I.~Antoniadis, J.~Rizos and K.~Tamvakis,
  Nucl.\ Phys.\  B {\bf 415}, 497 (1994)
  [arXiv:hep-th/9305025].


\bibitem{Brustein:1999yq}
  R.~Brustein and R.~Madden,
  JHEP {\bf 9907}, 006 (1999)
  [arXiv:hep-th/9901044].

\bibitem{Caprini:2003vc}
  C.~Caprini, R.~Durrer and T.~Kahniashvili,
  Phys.\ Rev.\  D {\bf 69}, 063006 (2004)
  [arXiv:astro-ph/0304556].

\bibitem{Kahniashvili:2005qi}
  T.~Kahniashvili, G.~Gogoberidze and B.~Ratra,
  Phys.\ Rev.\ Lett.\  {\bf 95}, 151301 (2005)
  [arXiv:astro-ph/0505628].

\bibitem{Spergel:2006hy}
  D.~N.~Spergel {\it et al.}  [WMAP Collaboration],
  arXiv:astro-ph/0603449.

\bibitem{Page:2006hz}
  L.~Page {\it et al.}  [WMAP Collaboration],
  arXiv:astro-ph/0603450.

\bibitem{Brustein:1998du}
  R.~Brustein and D.~H.~Oaknin,
  Phys.\ Rev.\ Lett.\  {\bf 82}, 2628 (1999)
  [arXiv:hep-ph/9809365].

\bibitem{Zaldarriaga:1996xe}
  M.~Zaldarriaga and U.~Seljak,
  Phys.\ Rev.\  D {\bf 55}, 1830 (1997)
  [arXiv:astro-ph/9609170].

\bibitem{Kamionkowski:1996ks}
  M.~Kamionkowski, A.~Kosowsky and A.~Stebbins,
  Phys.\ Rev.\  D {\bf 55}, 7368 (1997)
  [arXiv:astro-ph/9611125].

\bibitem{Hu:1997hp}
  W.~Hu and M.~J.~White,
  Phys.\ Rev.\  D {\bf 56}, 596 (1997)
  [arXiv:astro-ph/9702170].

\bibitem{Feng:2006dp}
  B.~Feng, M.~Li, J.~Q.~Xia, X.~Chen and X.~Zhang,
  Phys.\ Rev.\ Lett.\  {\bf 96}, 221302 (2006)
  [arXiv:astro-ph/0601095].

\bibitem{Liu:2006uh}
  G.~C.~Liu, S.~Lee and K.~W.~Ng,
  Phys.\ Rev.\ Lett.\  {\bf 97}, 161303 (2006)
  [arXiv:astro-ph/0606248].
  
\bibitem{Cabella:2007br}
  P.~Cabella, P.~Natoli and J.~Silk,
  arXiv:0705.0810 [astro-ph].

\bibitem{Scannapieco:1997mt}
  E.~S.~Scannapieco and P.~G.~Ferreira,
  Phys.\ Rev.\  D {\bf 56}, 7493 (1997)
  [arXiv:astro-ph/9707115].

\bibitem{Kosowsky:1996yc}
  A.~Kosowsky and A.~Loeb,
  Astrophys.\ J.\  {\bf 469}, 1 (1996)
  [arXiv:astro-ph/9601055].


\bibitem{Pritchard05}
  J.~R.~Pritchard and M.~Kamionkowski,
  Annals Phys.\  {\bf 318}, 2 (2005)
  [arXiv:astro-ph/0412581].
  
\bibitem{Zhang:2005nv}
  Y.~Zhang, W.~Zhao, T.~Xia and Y.~Yuan,
  Phys.\ Rev.\  D {\bf 74}, 083006 (2006)
  [arXiv:astro-ph/0508345].


\bibitem{Lewis:1999bs}
  A.~Lewis, A.~Challinor and A.~Lasenby,
  Astrophys.\ J.\  {\bf 538}, 473 (2000)
  [arXiv:astro-ph/9911177].


\bibitem{Fan:2006dp}
  X.~H.~Fan, C.~L.~Carilli and B.~Keating,
  Ann.\ Rev.\ Astron.\ Astrophys.\  {\bf 44}, 415 (2006)
  [arXiv:astro-ph/0602375]. 


\bibitem{Basko80} M.~M.~Basko and A.~G.Polnarev,
  Mon.\ Not.\ Roy.\ Astron.\ Soc.\ {\bf 191}, 207 (1980).

\bibitem{Zaldarriaga:1996ke}
  M.~Zaldarriaga,
  Phys.\ Rev.\  D {\bf 55}, 1822 (1997)
  [arXiv:astro-ph/9608050].

\bibitem{Lewis:2006fu}
  A.~Lewis and A.~Challinor,
  Phys.\ Rept.\  {\bf 429}, 1 (2006)
  [arXiv:astro-ph/0601594].

\bibitem{Zaldarriaga:1998ar}
  M.~Zaldarriaga and U.~Seljak,
  Phys.\ Rev.\  D {\bf 58}, 023003 (1998)
  [arXiv:astro-ph/9803150].

\bibitem{Planck:2006uk}
  Planck Collaboration,
  ``Planck: The scientific programme,''
  arXiv:astro-ph/0604069.




\bibitem{Lewis:2002ah}
  A.~Lewis and S.~Bridle,
  Phys.\ Rev.\  D {\bf 66}, 103511 (2002)
  [arXiv:astro-ph/0205436].

\bibitem{Bond:1998qg}
  J.~R.~Bond, A.~H.~Jaffe and L.~E.~Knox,
  Astrophys.\ J.\  {\bf 533}, 19 (2000)
  [arXiv:astro-ph/9808264].

\bibitem{erotto:2006rj}
    L.~Perotto, J.~Lesgourgues, S.~Hannestad, H.~Tu, Y.Y.Y~Wong, 
    JCAP {\bf10}, 013 (2006)
  [arXiv:astro-ph/0606227].


\bibitem{Seto:2006hf}
  N.~Seto,
  Phys.\ Rev.\ Lett.\  {\bf 97}, 151101 (2006)
  [arXiv:astro-ph/0609504].

\bibitem{Seto:2006dz}
  N.~Seto,
  Phys.\ Rev.\  D {\bf 75}, 061302 (2007)
  [arXiv:astro-ph/0609633].

\bibitem{Seto:2006ip}
  N.~Seto and A.~Taruya,
  Phys.\ Rev.\ Lett.\ in press (2007)

\bibitem{Seljak:1995ve}
  U.~Seljak,
  Astrophys.\ J.\  {\bf 463}, 1 (1996)
  [arXiv:astro-ph/9505109].

\bibitem{Lin:2004xy}
    Y.T.~Lin, D.~Wandelt,
  Astropart.\ Phys.\ {\bf 25}, 151 (2006)
  [arXiv:astro-ph/0409734].

\end{thebibliography}
\end{document}